\documentclass[aps,twocolumn,nofootinbib,tightenlines,showpacs,%
showkeys,floatfix,tightenlines,preprintnumbers]{revtex4}

\usepackage{epsfig,bm,feynmf}

\usepackage{graphics}

\usepackage[normalem]{ulem}  % \sout{old text} for strikeout
\usepackage[dvips]{color} % For blue in-text comments and additions

\begin{document}

%\preprint{hep-ph/0402135}

%%%%%%%%%%%%%%%%%%%%% Title %%%%%%%%%%%%%%%%%%%%%%

\title{$R_{AA}$ of $J/\psi$ near mid-rapidity in heavy ion collisions at $\sqrt{s_{NN}}=200$ GeV}

%%%%%%%%%%%%%%%%%%%% Authors %%%%%%%%%%%%%%%%%%%%%

\author{Taesoo Song}%
\email{songtsoo@yonsei.ac.kr}

\author{Woosung Park}%
\email{diracdelta@hanmail.net}

\author{Su Houng Lee}%
\email{suhoung@phya.yonsei.ac.kr}

\affiliation{Institute of Physics and Applied Physics, Yonsei
University, Seoul 120-749, Korea}

%%%%%%%%%%%%%%%%%%%% Abstract %%%%%%%%%%%%%%%%%%%%%

\begin{abstract}
We build up a model to reproduce the experimentally measured $R_{AA}$ of $J/\psi$ near midrapidty in Au+Au collision at $\sqrt{s_{NN}}=200$  GeV.  The model takes into account the $J/\psi$ suppression from the quark-gluon plasma and hadron gas as well as the nuclear absorption of primordial charmonia and the
regeneration effects at the hadronization stage, and hence is a generalization of the two component model introduced by Grandchamp and Rapp.  The improvements in this work are twofold; the addition of  the initial local temperature profile  and a consistent use of QCD NLO formula  for both the dissociation cross section in the hadron gas and the thermal decay widths in the quark-gluon plasma for the charmonium states.
The initial local temperature profile is determined from the assumption
that the local entropy density is proportional to a formula involving the number densities  of the number of participants and of the  binary collisions that reproduces the multiplicities of charged particles at
chemical freeze-out.  The initial local temperature profile brings about a kink in the $R_{AA}$ curve due to the initial melting of $J/\psi$.
%The volume of fireball at
%initial stage is divided to two regions, where local temperature is
%over the melting point of $J/\psi$ and where that is not. This
%scheme brings about a kink on $R_{AA}$ curve which seems to actually
%exist in experimental data.
%The average temperature of the fireball
%is determined by using thermal masses of quarks and gluon extracted
%from lattice QCD, and chemical potentials from the conservation of
%each flavor. The effective decay widths of charmonia both in
%quark-gluon plasma and in hadron gas are calculated up to the next
%to leading order in perturbative QCD with a proper value of coupling
%constant.
The initially formed fireball, composed of weakly interacting quarks and gluons with thermal masses that are extracted from lattice QCD, follows an isentropic expansion with cylindrical symmetry.   The fit reproduces well the Au+Au as well as the Cu+Cu data.  The same method is applied to predict the $R_{AA}$ expected from the Pb+Pb collision at LHC energy.
\end{abstract}

 \keywords{}

\maketitle

%%%%%%%%%%%%%%%%%%%% Text %%%%%%%%%%%%%%%%%%%%%

\section{Introduction}
Ever since the seminal work by Matsui and Satz \cite{Matsui:1986dk},
$J/\psi$  has been investigated for a long time as a diagnostic tool
to probe the properties of hot nuclear matter created in the early
stages of a relativistic heavy ion collision. The original claim
stated that $J/\psi$ cannot be formed in the quark-gluon plasma(QGP) due
to color screening between the charm and anti-charm quark and that
such mechanism will lead to the suppression of $J/\psi$ production
in a heavy ion collision if the QGP is formed.  However, the
situation was found to be more involved as recent
lattice QCD results showed that $J/\psi$ might survive past the
critical temperature ($T_c)$ and dissolve only at a higher
temperature ($T_d \sim 2 T_c$)\cite{Hatsuda04,Datta04}.  If so, it
is important to know the detailed temperature dependencies of the
properties of the $J/\psi$ above $T_c$, as  large thermal width for example
might still lead to a very small survival rate of the $J/\psi$\cite{Mocsy:2007jz,Morita:2007pt,Lee:2008xp}.
Moreover, if the quark-gluon plasma is formed and the number of
charm quarks are large, there is an additional production mechanism that has to be estimated and  comes from the formation of $J/\psi$  through the regeneration of charm and anti-charm that are well described by statistical approaches\cite{Gazdzicki:1999rk,BraunMunzinger:2000px,Gorenstein:2000ck,Thews:2000rj}. A successful phenomenological model
that is able to explain the $J/\psi$ signals coming from p+p to
heavy ion collisions at high energy should consistently include all
the above mentioned ingredients.

The commonly measured and calculated nuclear modification factor $R_{AA}$ shows  whether the $J/\psi$ is suppressed or enhanced in the heavy ion collision \cite{Back:2004je}. For instance, if $R_{AA}=1$, the number of produced $J/\psi$ in A+A collision is equal to  the number of produced $J/\psi$ in p+p collision at the same energy, times the number of binary collisions of nucleons from  the two colliding nuclei. If $R_{AA}<1 (>1)$, the number of $J/\psi$ dissolved in hot nuclear matter is larger (smaller) than that regenerated  from the quark-gluon plasma at the hadronization point.  Presently $R_{AA}$ of $J/\psi$ is below unity up to the maximum energy of RHIC. \cite{Adare:2006ns}.

Phenomenological models have been developed to describe $R_{AA}$ of $J/\psi$ from SPS to RHIC, and to predict the upcoming LHC heavy ion data.   One of the successful models to explain the heavy ion data is the thermal model \cite{Andronic:2006ky}.  Here the  $J/\psi$ production follows the statistical description at the chemical freeze out point, and thus can be thought of as being  regenerated from the  quark-gluon plasma at the hadronization point ($\sim T_c$).
In the so called two-component
model by Grandchamp and Rapp(GR) \cite{Grandchamp:2002wp}, one additionally takes into account the fact that the $J/\psi$ survives past $T_c$ and dissociates only at a higher temperature $T_d$ so that some of $J/\psi$ produced at initial stage survives the high temperature phase and reaches the final stage; these together with those regenerated from the quark-gluon plasma at the hadronization point constitutes the two components of the observed $J/\psi$  \cite{Zhao:2007hh}.
In a third model the continuous dissociation and regeneration of $J/\psi$  is calculated using the Boltzman transport equation from $T_d$ to $T_c$ evolved using the hydrodynamics equation  \cite{Yan:2006ve}.

Crucial quantities in the above models are the $T_d$'s for the charmonia; these are the melting temperatures for the Deby screened Coulomb type of charmonium  bound states above $T_c$.  While, initial calculations seemed to suggest that  $T_d \sim 2T_c$ for $J/\psi$ and lower for the excited states\cite{Hatsuda04,Datta04,Wong04}, recent studies claim that these could be lower due to the large thermal width the states acquire \cite{Mocsy:2007jz,Petreczky:2010yn}.
The current experimental error for  $R_{AA}$  of $J/\psi$ as a function of the  number of participants are still too large to draw any definite conclusions about these different temperatures.  However, one finds interesting characteristics if one takes  the central values of the experimental data for the curve of Au+Au collision at $\sqrt{s_{NN}}=200$  GeV; there seems to be two
sudden drops \cite{Adare:2006ns}. One occurs initially at a very small number of
participants and the other somewhere  between 170 and 210.  Assuming that the initial temperatures of the fireball is correlated to the number of participants, the two drops might be related to two distinct initial temperatures.  Indeed, as we will show in our two component model, the first sudden drop at
small number of participants can be related to the dissociation of
excited charmonia such as $\chi_c$ and $\psi'$; the initial dissociation of these states will suppress the expected 30 to 40\% feedback production
of $J/\psi$ from its excited states.  The second sudden drop is found to be related to the initial dissociation of $J/\psi$.  In geometrical terms, the first drop occurs when there appears regions where the initial temperature is larger than the $T_d$ of the excited charmonia, and the second drop when it is larger than $T_d$ of  $J/\psi$. Moreover, if the thermal decay width of $J/\psi$ is not so large, these effects become prominent.

In this work,  we generalize the two-component model to take into account the  initial hot region mentioned above, and calculate the $R_{AA}$ of $J/\psi$ at  midrapidity in Au+Au collision at $\sqrt{s_{NN}}=200$ GeV.
We estimate the initial local temperature with the assumption that local entropy density is simply the linear combination of number density of the number of  participants and of binary collision, a formula that well reproduce the multiplicities of charged particles at chemical freeze-out stage in the scenario of isentropically expanding fireball.
The average temperature of the fireball is obtained from equating the entropy density to that of  the quark-gluon plasma calculated using a non interacting gas of  quarks and gluons with effective thermal masses extracted from lattice QCD calculation for the  energy density and pressure \cite{Levai:1997yx}.
%In quasiparticle model, strongly interacting massless particles can be described %%with noninteracting massive particles which result in the same thermal quantities %such as energy density and pressure.
The baryon chemical potential is obtained from the ratio of the proton and the antiproton, and other chemical potentials from the conservation of respective  flavors.

In the two-component model, the primordial charmonia, which have not evolved to the final  charmonia yet, first undergo nuclear absorption.  Then after they form into the initially produced charmonia, they  undergo through thermal decay in the quark-gluon plasma,  and then hadronic decay in the hadron gas until the chemical freeze-out stage.   On the other hand, the regenerated charmonia only go through the hadronic decay in the hadron gas.
Consequently, another important quantity in the model is the thermal widths of charmonia in the quark-gluon plasma and in hadron gas.
We improve previous calculations by using the results obtained using perturbative QCD up to the next to leading order (NLO) in the coupling constant\cite{Song:2005yd}.
In our work, this coupling constant is not a free parameter, because it is related to the screening mass of quark-gluon plasma and sequential melting temperatures of charmonia. The details is mentioned in chapter VII.
The relaxation factor of charm quarks in quark-gluon plasma is calculated up to leading order in perturbative QCD; this factor determines the fraction of  thermalized charm quarks and used to estimate the regeneration of charm quarks.

The paper is organized as follows.  In section II, we discuss nuclear absorption and introduce the necessary concepts.  In section III, we introduce the initial local temperature profile.  In section IV, we show how the thermodynamic parameters are determined. In section V and VI we respectively discuss the thermal decay and regeneration of charmonia.  In the last two sections, we give the results and conclusions.  In the appendix, we summarize the perturbative NLO formula for the dissociation cross section of charmonium.

\section{nuclear absorption}
Within the Glauber model, heavy ion collision can be described with collisions of the nucleons.
The model has two important scales, the number of participants and the number of binary collisions.
Participants mean the nucleons in both colliding nuclei that go through inelastic scattering at least once. The rest are called spectators. Usually the amount of bulk matter created from heavy ion collision is proportional to the number of participants. The number of binary collisions counts only primary collisions of two nucleons. Usually hard particle such as the charmonium is produced through primary collision, because it requires large energy transfer; hence, its production number is proportional to the number of binary collisions.
The two numbers are calculated in Glauber model as follows \cite{Antinori:2000ph}:

\begin{eqnarray}
N_{part}(\vec{b})&=&A\int d^2 s T_A(\vec{s})\bigg[ 1-\{ 1-T_B(\vec{b}-\vec{s})\sigma_{in}\}^B \bigg] \nonumber\\
&+& B\int d^2s T_B(\vec{b}-\vec{s})\bigg[1-\{1-T_A(\vec{s})\sigma_{in}\}^A\bigg] \nonumber\\
N_{coll}(\vec{b})&=&\sigma_{in}AB \int d^2 s T_A(\vec{s}) T_B(\vec{b}-\vec{s})\nonumber\\
&\equiv& \sigma_{in}AB~ T_{AB}(\vec{b}),
\label{two-scales}
\end{eqnarray}
where $\vec{b}$ is the impact parameter of the two colliding heavy nuclei, and A, B are their mass numbers.
$\sigma_{in}$ is the inelastic cross section of two nucleons,
which is about $42\ mb$ at $\sqrt{s_{NN}}=200$ GeV \cite{Back:2004je}. $T_{A(B)}$ is called the thickness function defined as
\begin{equation}
T_{A(B)}(\vec{s})=\int_{-\infty}^\infty dz \rho_{A(B)}(\vec{s},z),
\end{equation}
and $T_{AB}$ is called the overlap function.
Here, $\rho_{A(B)}(\vec{r})$ is the distribution function of one nucleon in nucleus A(B), for which we use the following Woods-Saxon model:
\begin{equation}
\rho(r)=\frac{\rho_o}{1+e^{(r-r_o)/C}}.
\label{Woods-Saxon}
\end{equation}
For the Au nucleus, $r_o=6.38\ {\rm fm}, C=0.535\ {\rm fm}$  \cite{De Jager:1987qc}.
The normalization factor $\rho_o$ is determined by the requirement $\int d^3r \rho(r)=1$.

The charm and anticharm quark pair produced through primary
collision is called the primordial charmonia before they form into
on shell states.  Such primordial charmonia can still be absorbed by
nucleons passing through it. The survival rate at transverse
position $\vec{s}$ from the nuclear absorption is

\begin{eqnarray}
S_{nuc}(\vec{b},\vec{s})=\frac{1}{T_{AB}(\vec{b})}\int dz dz' \rho_A(\vec{s},z)\rho_B(\vec{b}-\vec{s},z')\nonumber\\
\times {\rm exp}\bigg\{ -(A-1)\int_z^\infty dz_A \rho_A (\vec{s},z_A)\sigma_{nuc}\bigg\}\nonumber\\
\times {\rm exp}\bigg\{ -(B-1)\int_{z'}^\infty dz_B \rho_B
(\vec{b}-\vec{s},z_B)\sigma_{nuc}\bigg\}.
\label{absorption}
\end{eqnarray}
Here, $\sigma_{nuc}$ is the absorption cross section of primordial
charmonia by a nucleon, and is about $1.5\ mb$ at
$\sqrt{s_{NN}}=200$ GeV \cite{Zhao:2007hh}. $z$ and $z'$ are the
positions where the primordial charmonium is created in beam
direction from the center of nucleus A and from the center of
nucleus B respectively. The second line in Eq.~(\ref{absorption}) is
the absorption factor by nucleus A, and the third by nucleus B. Mass
number A and B in the exponents of the second and third line are
both subtracted by 1, because one nucleon of nucleus A and one
nucleon of nucleus B take part in producing a charm pair, and cannot
take part in absorbing the pair.

\section{initial local temperature}

To build up a model for the suppression, we will closely follow the
space time picture of nucleus nucleus collision generally accepted
from a hydrodynamic simulations\cite{Morita02}. After the heavy ions
pass through each other, hot nuclear matter is created with  small
net baryon density between the two receding nuclei. The hot matter
is expected to  thermalize early; while the exact thermalization time could be controversial, we will just take the typical time accepted in hydrodynamic simulation given as $\tau_0=0.6 $fm/c\cite{Hirano:2001eu,Gunji:2007uy}. 

The total entropy
produced in heavy ion collision is estimated  from the multiplicity
of produced particles. The multiplicity of charged particles near
midrapidity was found to be well described under the assumption that
it is proportional to a linear combination of the number of
participants and the number of binary collisions. Because both the
multiplicity of particles and the entropy are extensive thermal
quantities, it is assumed that the entropy per pseudorapidity is
also proportional to the same linear combination of number of
participants and binary collisions as the particle multiplicity.
\begin{equation}
\frac{dS}{d\eta}=c \frac{dM}{d\eta}= 30.3 \bigg\{(1-x)\frac{N_{part}}{2}+xN_{coll}\bigg\}.
\label{entropy}
\end{equation}

In the linear combination for multiplicities of charged particles, $x$
is 0.09 at $\sqrt{s_{NN}}=130$ GeV and 0.11 at
$\sqrt{s_{NN}}=200$ GeV; the same values are used for entropy
\cite{Kharzeev:2000ph,Back:2002uc}. The overall factor of 30.3 is
determined so that multiplicities of charged particles are well
reproduced under the condition that the ratio between entropy and
multiplicity is determined from the statistical model at the
chemical freeze out point; $c=S/M=s/{n_M}$, where $S,M$ $(s,n_M)$
are the total entropy and multiplicity (density) respectively. The
upper figure of Fig. \ref{multiplicity} shows that Eq.
(\ref{entropy}), which is the solid line, reproduces the data well,
when isentropic expansion of hot nuclear matter is assumed.

Dividing Eq. (\ref{entropy}) by the volume, entropy density per
pseudorapidity has the form

\begin{equation}
\frac{dS}{V d\eta}\equiv \frac{ds}{d\eta}=30.3\bigg\{(1-x)\frac{n_{part}}{2}+xn_{coll}\bigg\},
\label{entroden}
\end{equation}
where $n_{part}$, $n_{coll}$ are volume densities of the number of participants and binary collisions respectively. These densities at thermalization time $\tau_o$, which is taken to be 0.6 $fm/c$ in this work, are given respectively as,
\begin{eqnarray}
n_{part}(\vec{s})&=&A T_A(\vec{s})\bigg[ 1-\{ 1-T_B(\vec{b}-\vec{s})\sigma_{in}\}^B \bigg]/\tau_o \nonumber\\
&+& B T_B(\vec{b}-\vec{s})\bigg[1-\{1-T_A(\vec{s})\sigma_{in}\}^A\bigg]/\tau_o \nonumber\\
n_{coll}(\vec{s})&=&\sigma_{in}AB T_A(\vec{s}) T_B(\vec{b}-\vec{s})/\tau_o,
\end{eqnarray}
where $\vec{s}$ is the position vector on the transverse plane, and homogeneous densities in beam direction is assumed. 

Near midrapidity, the rapidity is approximated to the longitudinal velocity,
\begin{equation}
y\equiv \frac{1}{2}\ln \frac{E+p_z}{E-p_z}\approx \beta_z,
\end{equation}
and is further assumed to be equal to the pseudorapidity. Supposing
that the rapidity interval of interest is $\Delta y$, the total
entropy within the interval is $(dS/d\eta) \Delta y$. Ignoring
longitudinal acceleration, longitudinal size of nuclear matter
within $\Delta y$ is $\tau_o \beta_z  \approx \tau_o \Delta y$ at
thermalization time $\tau_o$. Therefore, entropy density at
thermalization time is

\begin{equation}
\frac{ds}{d\eta}\Delta \eta \approx \frac{dS}{d\eta}\frac{\Delta y}{A \tau_o \Delta y }=\frac{1}{A\tau_o}\frac{dS}{d\eta},
\end{equation}
where A is the transverse area.

The equation of state is required to extract the temperature from the entropy density. In the quasi-particle picture, the strongly interacting massless particles are replaced by the noninteracting massive particles, whose properties are determined  from the condition that they reproduce the thermal quantities such as energy density and pressure of the strongly interacting massless particles. The thermal quantities of strongly interacting quarks and gluons are simulated by lattice QCD.  In this work, we will use the parametrization for the effective thermal masses of quark and gluon extracted by  Levai and Heinz from lattice QCD data for pure gauge, for $N_f=2$ and for $N_f=4$ \cite{Levai:1997yx}. They fix the degrees of freedom of quarks and gluon to those in the vacuum, and use the forms of LO perturbative QCD for their masses, parameterizing the the nonperturbative effects into a temperature dependent coupling $g(T)$:

\begin{eqnarray}
m_g^2&=&\bigg(\frac{N_c}{3}+\frac{N_f}{g}\bigg)\frac{g^2(T)T^2}{2}, \nonumber\\
m_q^2&=&\frac{g^2(T)T^2}{3},
\end{eqnarray}
where
\begin{eqnarray}
g^2(T)&=&\frac{48\pi^2}{(11N_c-2N_f)\ln F^2(T,T_c,\Lambda)}, \label{lhcoupling} \\
F(T,T_c,\Lambda)&=&\frac{18}{18.4e^{(T/T_c)^2/2}+1}\frac{T}{T_c}\frac{T_c}{\Lambda}.
\end{eqnarray}
$T_c=260$, 140, 170  MeV and $T_c/\Lambda=1.03$, 1.03, 1.05 for pure
gauge, $N_f=2$ and $N_f=4$ cases respectively. Here we choose the
case of $N_f=4$ for thermal masses, because its critical temperature
is more reasonable than that of $N_f=2$.

\begin{figure}
\centerline{
\includegraphics[width=5.5cm, angle=270]{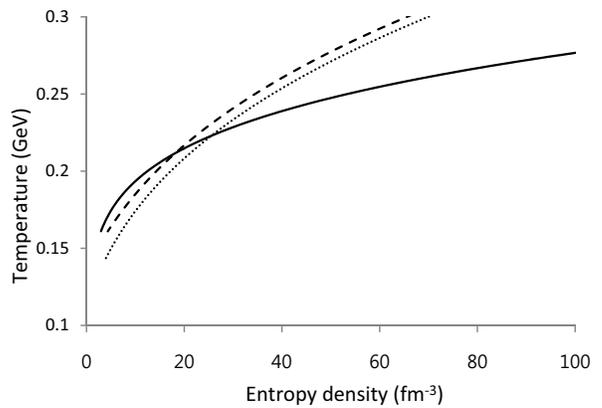}}
\caption{Entropy density vs. temperature in the hadron gas (solid line) and in the quark-gluon plasma with parameters extracted from lattice QCD with two flavors (dotted line) and with four flavors (dashed line) } \label{tem-entropy}
\end{figure}

With these effective
thermal masses and the equation of state obtained by assuming a gas of weakly interacting quarks and gluons, the entropy
density is related to temperature as given below,
\begin{eqnarray}
s= -\frac{1}{V}\frac{\partial \Phi}{\partial
T}\bigg|_{V,\mu}=\frac{1}{6\pi^2T^2}\sum_i \int d k k^4 \frac{
e^{E_i/T}}{(e^{E_i/T}\pm 1)^2}
\label{entropy-den}
\end{eqnarray} 
The
entropy density of quark-gluon plasma has  a term proportional to the
derivative of the squared thermal mass with respect to temperature,
because thermal masses depend on temperature. But the term is to be
canceled by the term induced from bag pressure to maintain
thermodynamic consistency \cite{Schneider:2001nf}. The summation $i$
includes gluon and quark of three flavors, although the lattice data
is extracted for $N_f=4$, we use the realistic number of flavors to
get a more realistic magnitude for the entropy. The signs in the
denominators are minus for gluon and plus for quarks. Dashed line
and dotted line in Fig. \ref{tem-entropy} show the entropy density as a function of the temperature in quark-gluon plasma respectively for parameters extracted from lattice QCD for $N_f=4$ and $N_f=2$.

\begin{figure}
\centerline{
\includegraphics[width=5.5cm, angle=270]{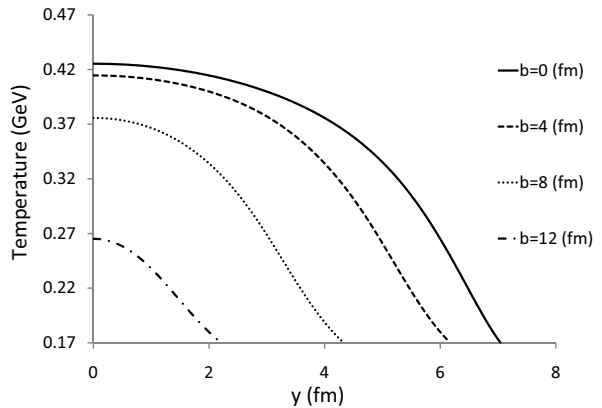}}
\caption{Profiles of temperature along y-axis at various impact parameters} \label{profile}
\end{figure}

\begin{figure}
\centerline{
\includegraphics[width=5.5cm, angle=270]{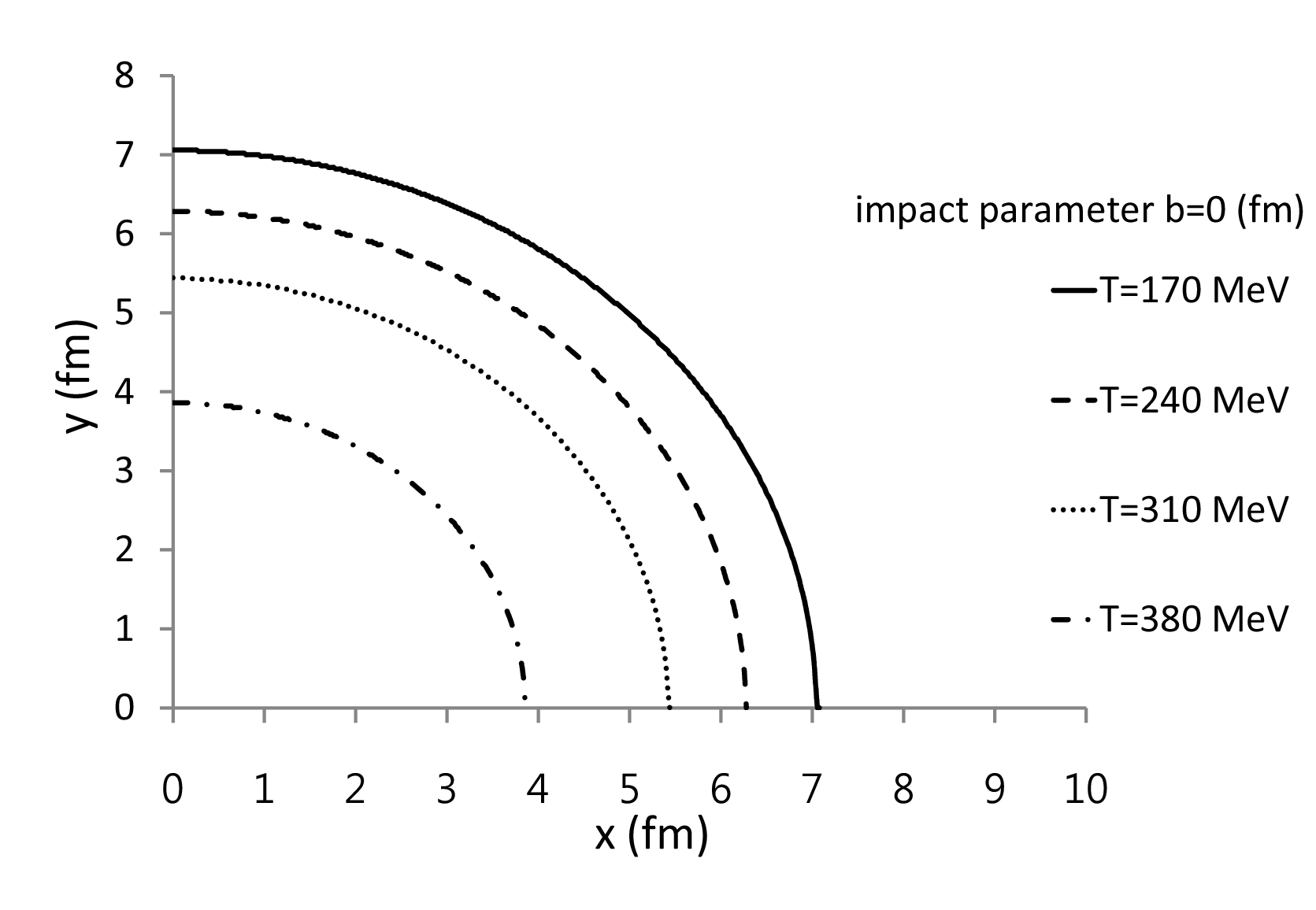}}
\centerline{
\includegraphics[width=5.5cm, angle=270]{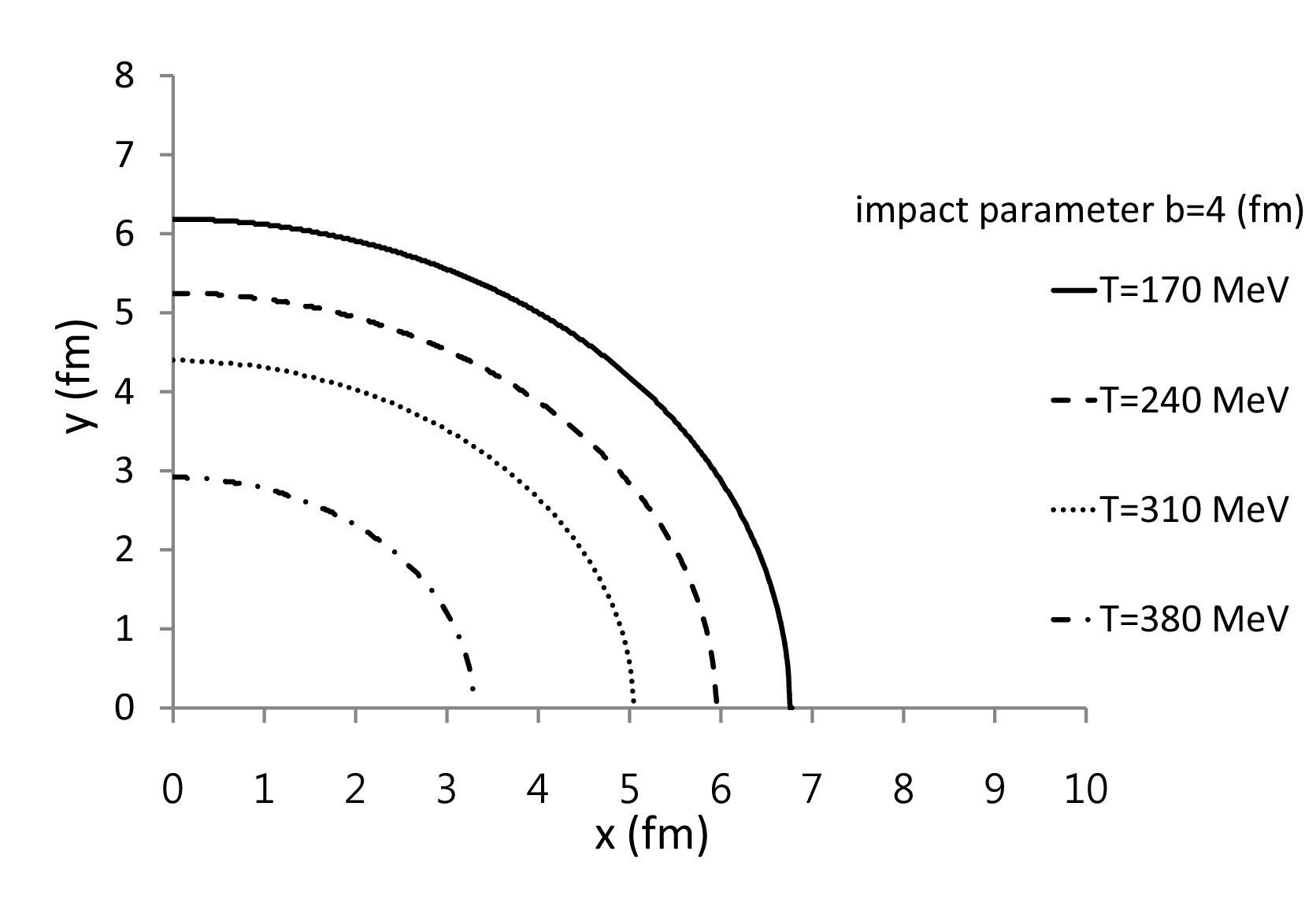}}
\centerline{
\includegraphics[width=5.5cm, angle=270]{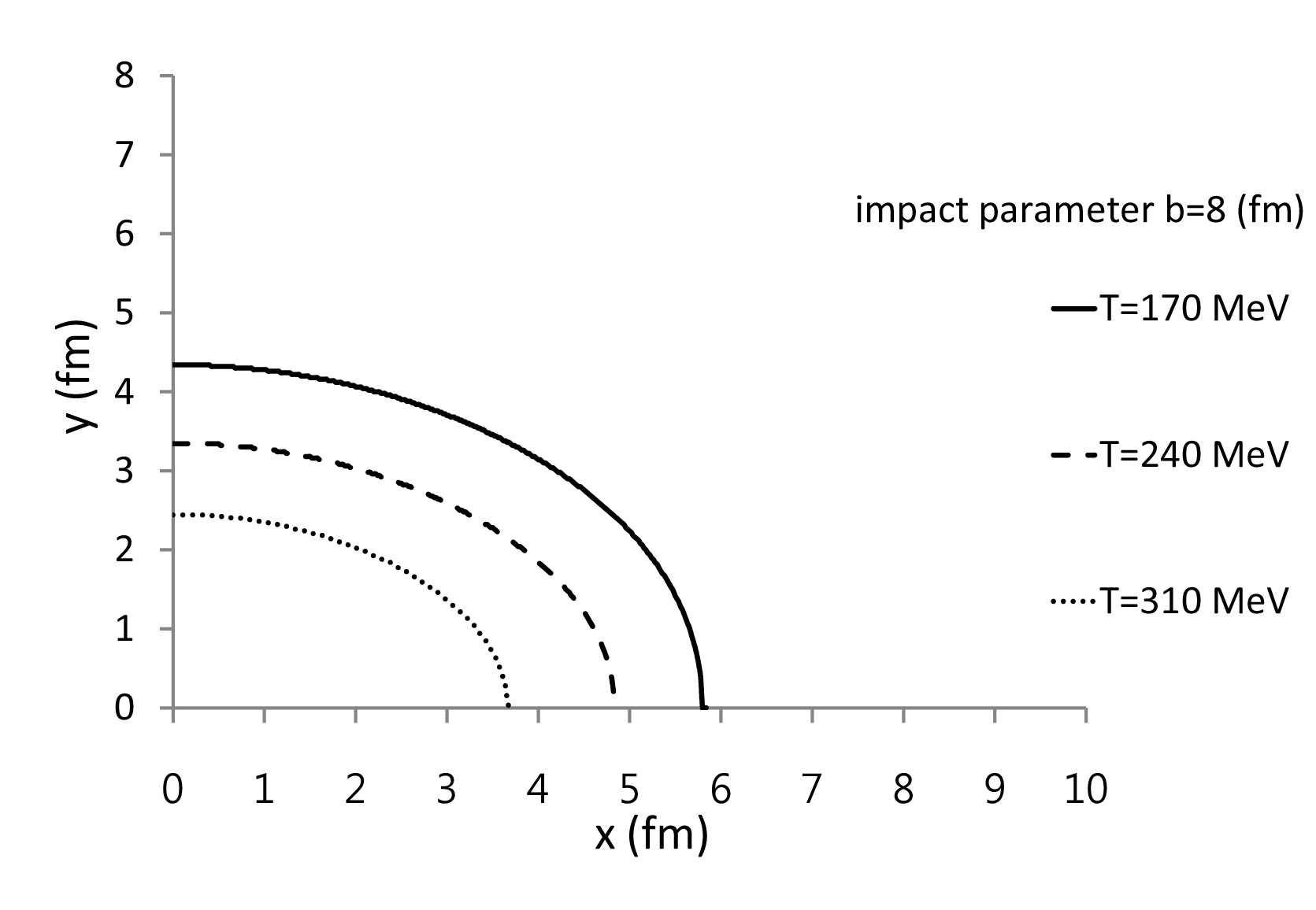}}
\caption{Isothermal lines on transverse plane at $\tau=\tau_o$ with impact parameters, $b=0, 4, 8\ {\rm fm}$}
\label{isothermal}
\end{figure}

From the entropy density of Eq. (\ref{entroden}) and the relation between entropy density and temperature given in Eq. (\ref{entropy-den}), the initial local temperature distribution at $\tau=\tau_o$ is obtained as given in Fig. \ref{profile} and \ref{isothermal}. y-axis is the direction of impact parameter and x-axis is perpendicular to beam axis and y-axis. Fig. \ref{profile} shows temperature profiles along the y-axis at various impact parameters. As we can see, the maximum temperature becomes lower as impact parameter increases. Fig. \ref{isothermal} shows isothermal lines on transverse plane. In head-on collision case, the transverse radius of quark-gluon plasma is about 7.0  fm at thermalization time. This size decreases as the collision becomes peripheral. More importantly, the equithermal line of $T=380~$ MeV  exists at $b=0,~4$ fm, but does not at $b=8$ fm. If the melting temperature of $J/\psi$ is 380 MeV, the region where $J/\psi$ cannot be formed exists at $b=0$ and at $b=4$ fm, but does not at $b=8$ fm, which, as will be seen later, leads to a sudden drop to $R_{AA}$ curve of $J/\psi$ as a function of the number of participants.

\section{thermal quantities in expanding hot nuclear matter}
The hot nuclear matter created through a heavy ion collision expands in time; assuming isentropic expansion, the entropy density and the temperature decreases along with it. Once the critical temperature is reached, the phase of the matter changes from quark-gluon plasma to the hadron gas. It is known that the phase transition is a crossover at small baryon chemical potential and that there is a critical point where the phase transition changes from a crossover to the first order as baryon chemical potential increases.  While the exact location of the point is not certain yet,  the phase transition is expected to be still at the rapid crossover region at RHIC and at LHC energies.

The statistical model is very successful in reproducing the observed particle ratios with two free parameters: the temperature and the baryon chemical potential at the chemical freeze-out stage. The other chemical potentials are obtained from the condition that each flavor must be conserved, which can be written as follows:
\begin{eqnarray}
V\sum_j n_j B_j &=& Z+N \nonumber\\
V\sum_j n_j I_{3j}&=& \frac{Z-N}{2}\nonumber\\
V\sum_j n_j S_j &=& 0.
\label{conservation}
\end{eqnarray}
Here $V$ is the volume of the hot nuclear matter of interest, $Z$ and $N$ the number of protons and neutrons in that volume respectively, and $j$ the constituents of the matter: the constituents are the quarks and gluons in the quark-gluon plasma, while they are the  mesons and baryons in the hadron gas.
For hadron gas, all mesons and baryons whose masses are less than 1.5 GeV and 2.0 GeV respectively are included in the sum. These upper limits for the hadron masses are taken such that inclusion of hadrons with masses higher than the upper bounds will not change the result; for example, if the upper limit of meson mass is set at 2.0 GeV, much more mesons are to be included, but their contribution is small.
$n_j$ is the number density of particle $j$ in the grandcanonical ensemble. $B_j$, $I_{3j}$ and $S_j$ are baryon number, the third component of isospin and strangeness of particle $j$ respectively.
%Note, that $Z+N$ is the net number of baryons in volume $V$.
%If $V$ is total volume of fireball, $Z+N$ will be the number of participants.

In our work, $V$ will represent the volume in the midrapidity region, which is defined by $|y|<0.35$.
To determine all the chemical potentials from Eq.~(\ref{conservation}), we have to know the two numbers $Z$ and $N$.
The two numbers are determined from two constraints: the first is  that the ratio $Z/N$ should be the same as in the original combined nuclei, and the second is that  the observed ratio of proton to antiproton within the rapidity is well  reproduced. The second constraint needs some more explanation.  Instead of solving Eq.~(\ref{conservation}) for each $N_{part}$, we will try to find the best fit for the experimentally observed  $p/\bar{p}$ ratio as a function of $N_{part}$ by assuming that $Z+N$ appearing in the right hand side of the first equation in Eq.~(\ref{conservation}) is equal to the total number of participants in the collision divided by a constant number; the constant number is found to be 20.5.

\begin{figure}
\centerline{
\includegraphics[width=5.5cm, angle=270]{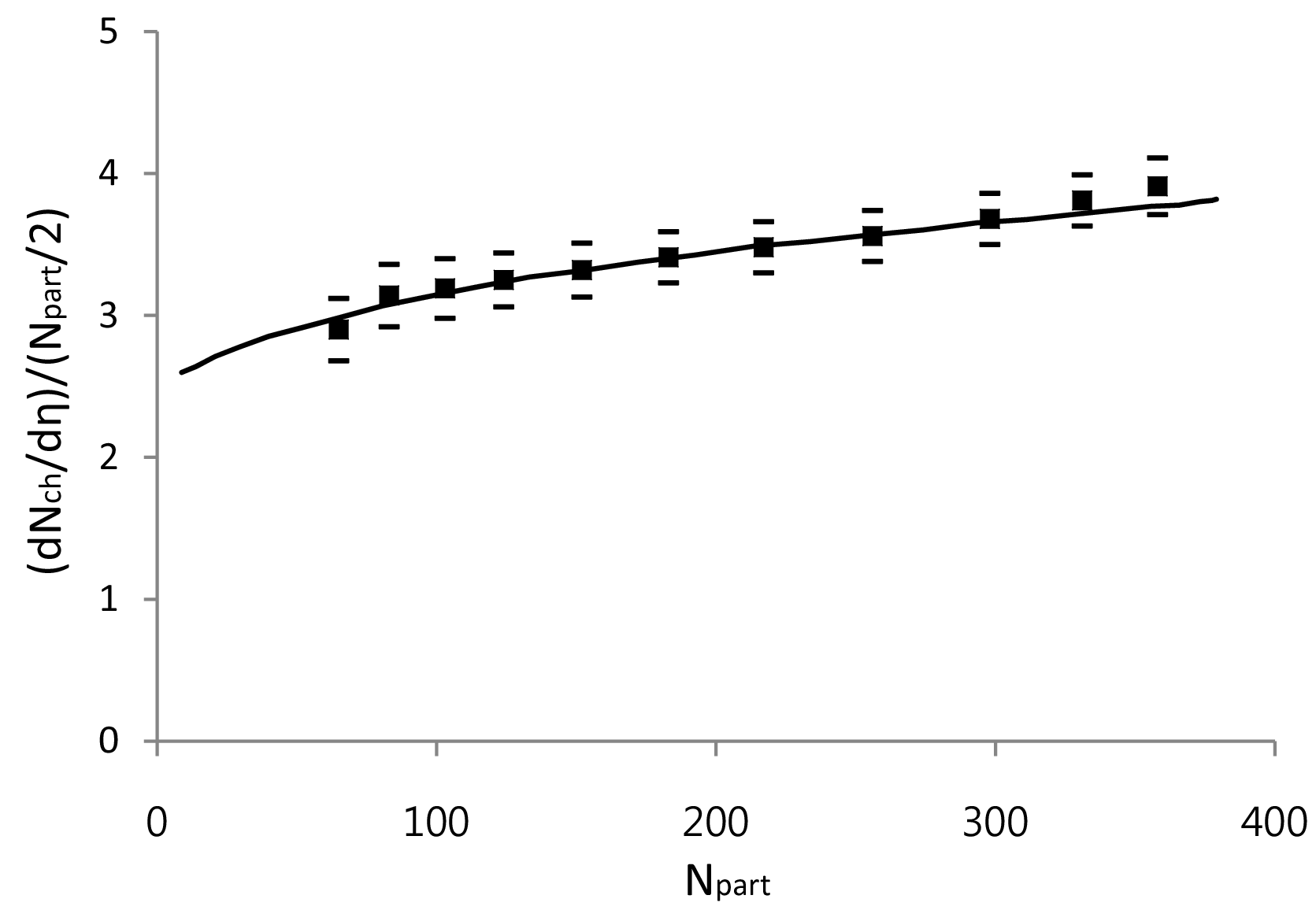}}
\centerline{
\includegraphics[width=5.5cm, angle=270]{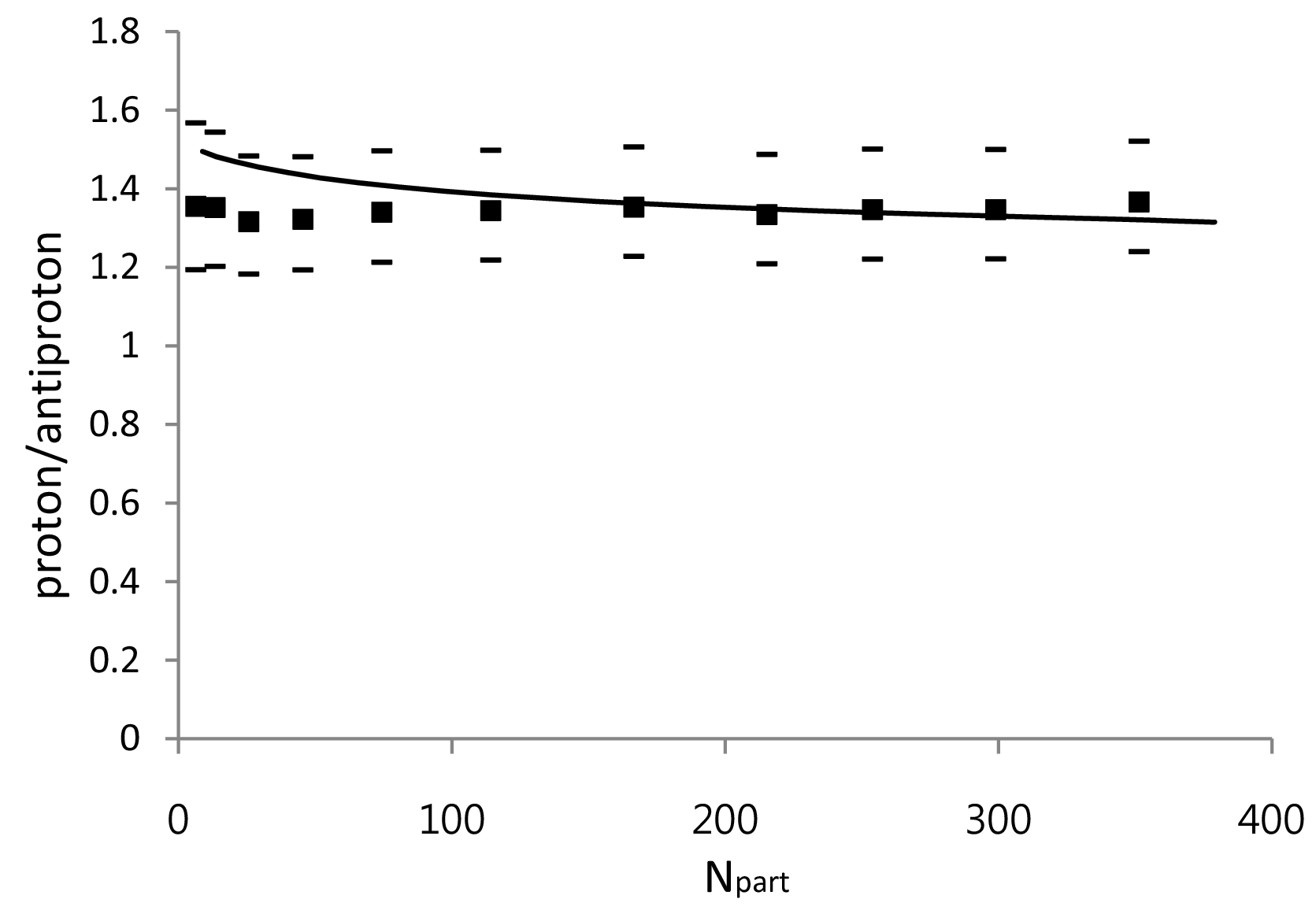}}
\caption{multiplicities of charged particles per pseudorapidity divided by a half number of participants (upper) and ratios of proton and antiproton (lower) near midrapidity at $\sqrt{s_{NN}}=200$ GeV} \label{multiplicity}
\end{figure}

The relation between $Z+N$ and the ratio of proton and antiproton is as following. If $Z+N$ is very large, the nuclear matter with volume $V$ will have much more baryons  than antibaryons and thus baryon chemical potential will be large and positive; in this case, the proton to antiproton ratio will be much larger than 1. On the other hand, if $Z+N$ is very small, the ratio will be close to 1. As one can see in the lower figure of Fig. \ref{multiplicity}, the best fit shown as the solid curve well reproduces the experimental data.

For the volume $V$, the following cylindrical form is used for simplicity:
\begin{equation}
V=2\beta(\tau_o +\tau)\pi \bigg(r_o+\frac{1}{2}a_{\bot}\tau^2 \bigg)^2.
\label{volume}
\end{equation}
The factor 2 is multiplied to take into account the  forward and the backward expansion in the beam direction.
$\beta$ is the longitudinal velocity of nuclear matter, which is approximately equal to the rapidity near midrapidity. $\tau_o$ is thermalization time, and $r_o$ is the initial radius of quark-gluon plasma on the  transverse plane.
As can be seen in Fig. \ref{isothermal}, the shape of quark-gluon plasma is not a circle but almond-like except for the head-on collision case.

In this work, the almond shape is transformed to a circle  with the same area. $a_{\bot}$ is the transverse acceleration of quark-gluon plasma, which is set to $0.1\ c^2/$ fm \cite{Zhao:2007hh}.
The acceleration is turned off, when the transverse velocity reaches 0.6 c.
The initial total entropy  of the quark-gluon plasma is calculated from a modified version of   Eq. (\ref{entropy}), where the $N_{part}$ and $N_{coll}$ are not the total number of participants and total number of binary collisions but are respectively the number of participants and the number of binary collisions within quark-gluon plasma.  Then, the entropy density is obtained by dividing the total entropy of the quark-gluon plasma by the initial volume as given in Eq. (\ref{volume}) at $\tau=\tau_0$.  Finally, the entropy density at a later invariant time is obtained by assuming the expansion to be isentropic and that the volume expands as given in Eq. (\ref{volume}).

With the chemical potentials obtained from Eq. (\ref{conservation}),
the entropy density determines temperature. The evolution of
temperature of the quark-gluon plasma is obtained by equating the
expression for the entropy density given in Eq. (\ref{entropy-den})
to the previously determined entropy density at a later time. In
contrast to the case for the quark-gluon plasma, it is assumed that
the  masses of all hadrons do not change at finite temperature.
%Only mesons whose masses are less than 1.5 $GeV$ and baryons whose masses are less than 2.0 $GeV$ are considered again for entropy density of hadron gas, for the contribution of heavier hadrons to entropy density is negligible at temperature of few hundred MeV.
As we discussed before, the order of the phase transition that we
probe at RHIC seems to be a strong cross over.  However, the
equation of states for the quark-gluon plasma and the hadron gas
that we use gives a 1st order transition.  Nevertheless, we will
still use these equation of states because they are simple,
physically intuitive and easy to manipulate.  Moreover, although at
the physical quark masses, the order of phase transition is a rapid
cross over, the change occurs at a very small temperature range so
that effectively we can approximate the transition with a simple 1st
order transition.

\begin{figure}
\centerline{
\includegraphics[width=5.5cm, angle=270]{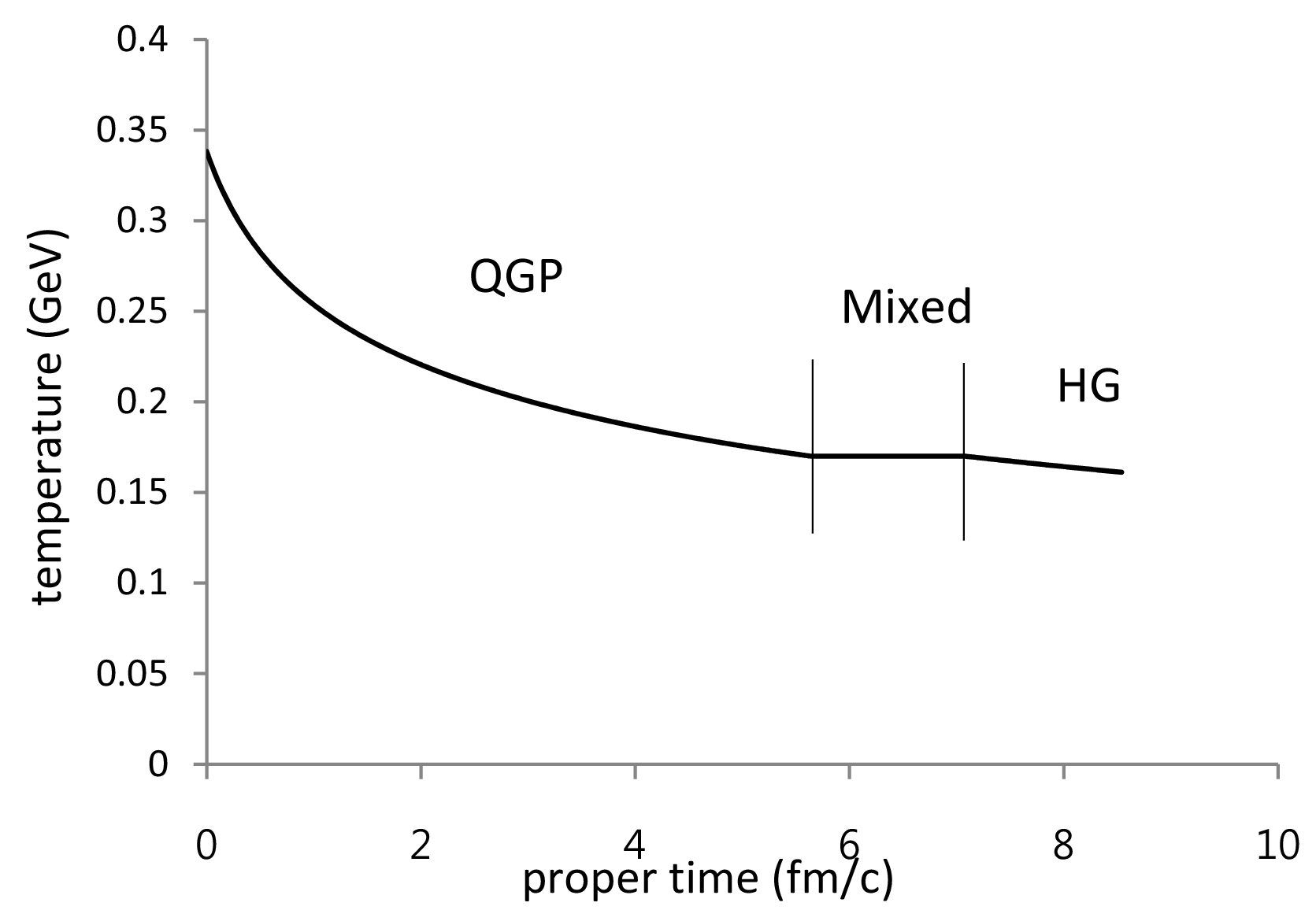}}
\centerline{
\includegraphics[width=5.5cm, angle=270]{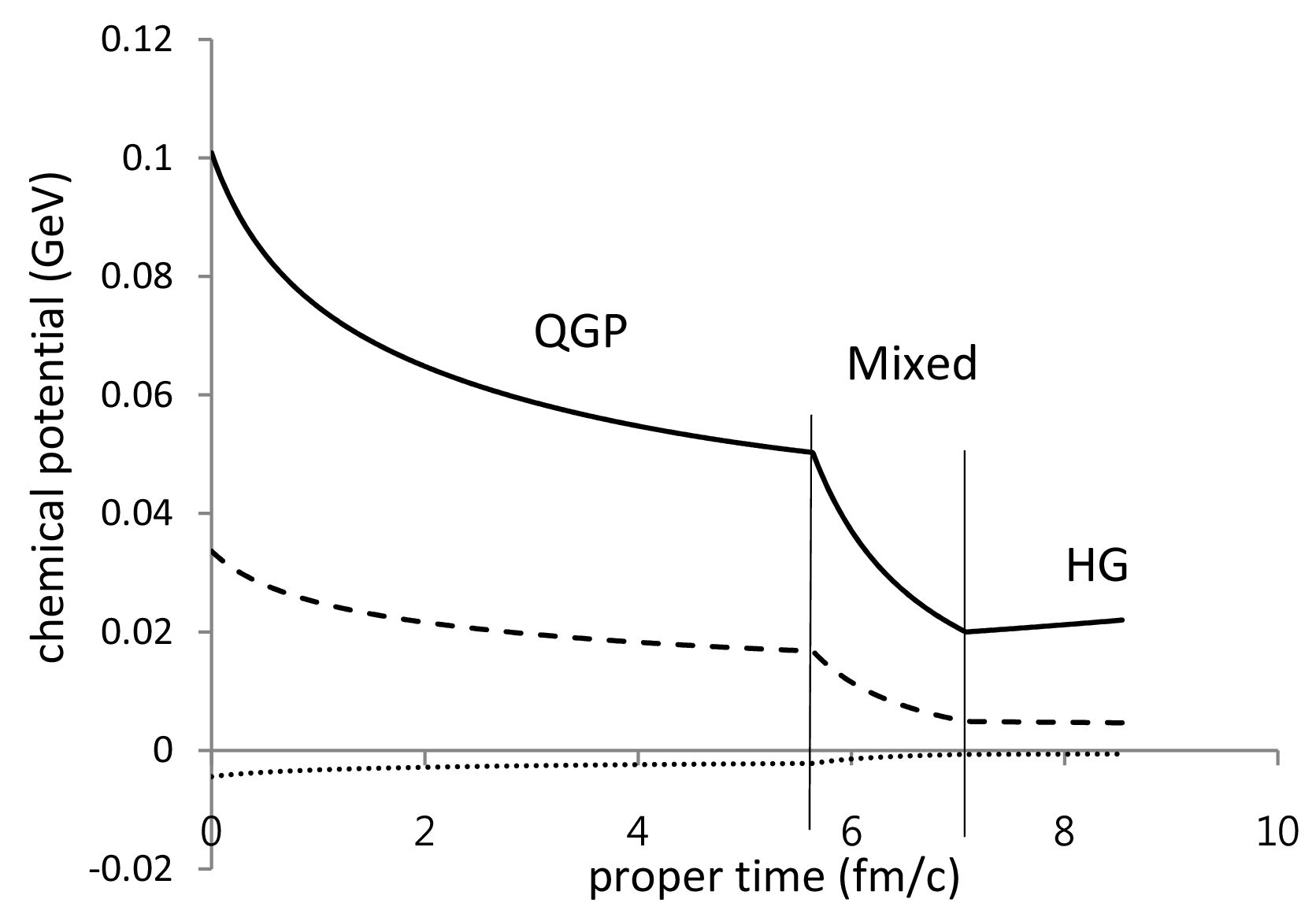}}
\caption{The time evolution of average temperature(upper figure) and chemical potentials(lower figure) of hot nuclear matter at $b=0\ $ fm. The solid line, dotted line, and dashed line in the lower figure are baryon, strangeness and isospin chemical potentials respectively.} \label{temp-potentials}
\end{figure}

The solid line in Fig. \ref{tem-entropy} shows the
correspondence between the entropy density to  the temperature in
the hadron gas. As can be seen in Fig. \ref{tem-entropy}, the
entropy density of the hadron gas is higher than that of quark-gluon
plasma at high temperature.  These curves cross at around
$T=210$ MeV below which the entropy density of hadron gas is lower
than that of quark-gluon plasma. But the curve of hadron gas and
curves of quark-gluon plasma are very close to each other, which means
that the phase transition from quark-gluon plasma to hadron gas is fast. As
hot nuclear matter expands, its entropy density and temperature
decreases along the curve of quark-gluon plasma in Fig.
\ref{tem-entropy}. At critical temperature, the nuclear matter
transfers from the curve of quark-gluon plasma to that of hadron
gas. If two curves meet at critical temperature and their
derivatives with respect to temperature are the same, the phase
transition is a crossover. If two curves meet but their derivatives
are different, the phase transition is second order. But two curves
in Fig. \ref{tem-entropy} do not meet at critical temperature so the
phase transition is first order. However, the phase transition takes
short time, because two curves are close at critical temperature.
This time interval is the period of mixed phase where two phases
coexist.   The upper
figure of Fig. \ref{temp-potentials} shows time dependence of the average temperatures of the
hot nuclear matter in head-on collision.
%The hot
%nuclear matter of here means nuclear matter whose initial
%temperature is over critical temperature, that is, nuclear matter
%within the solid line in Fig. \ref{isothermal}.

The lifetime of
quark-gluon plasma phase in head-on Au+Au collision at
$\sqrt{s_{NN}}=200$ GeV is 5.63 fm/c. Mixed phase lasts for 1.44
fm/c. Temperature for chemical freeze-out is set at 161 MeV, as motivated by thermal models. After chemical freeze-out, the number of
particles of each species is assumed to remain constant.
Multiplicities of charged particles and proton to antiproton ratio are all calculated at
this temperature. Hadron gas phase lasts for 1.52 fm/c to chemical
freeze-out.

Charmonia produced outside hot region where initial temperature is higher than dissociation temperature of charmonia go through thermal decay caused by interactions with surrounding particles, which are the quarks and the gluons in the  quark-gluon plasma, and the hadrons in the hadron gas or both in the mixed phase. Once thermal quantities of hot nuclear matter are given as functions of time, one needs to know the properties of charmonia in hot nuclear matter in order to calculate how many charmonia survive the thermal dissociation.

\section{Thermal decay of charmonia}

The thermal decay width of charmonia at temperature $T$ can be calculated using the following factorization formula \cite{Song:2007gm}:

\begin{eqnarray}
\Gamma(T)=\sum_i d_i \int
\frac{d^3k}{(2\pi)^3}v_{rel}(k)n_i(k,T) \sigma_i^{diss}(k,T),
\label{width}
\end{eqnarray}
where $i$ repents the quark or the gluon in the quark-gluon plasma, or the  baryons and the meson in the hadron gas. $d_i$ is the degeneracy factor of particle $i$, $n_i$ the number density of particle $i$ in the grand canonical ensemble, $v_{rel}$ the relative velocity between the charmonium and the particle $i$, and $\sigma_i^{diss}$ the dissociation cross section of charmonium by particle $i$. It is assumed that thermal width in the mixed phase is a linear combination of contributions  from the quark-gluon plasma and from the hadron gas as given in the following form:

\begin{equation}
\Gamma=f\Gamma^{HG}+(1-f)\Gamma^{QGP},
\end{equation}
where $f$ is the fraction of the hadron gas in the mixed phase.

\subsection{dissociation cross section}

The crucial quantity in Eq. (\ref{width}) is the dissociation cross
section. As emphasized before, one of the improvements in our work over the previous two component model calculations is the
consistent application of the NLO perturbative formula to calculate the dissociation of charmonia both in the quark-gluon plasma \cite{Park:2007zza} and in the hadron gas \cite{Song:2005yd}.   In fact, it is difficult to describe the dissociation of charmonia
both in the quark-gluon plasma and in the hadron gas with the same approach.
As an example, GR in ref.\cite{Grandchamp:2002wp} uses a meson exchange model for estimating the dissociation of
charmonia in the hadron gas and a quasifree particle approximation for charm quark inside the $J/\psi$ in the quark-gluon plasma. Binding energies of
charmonia become very small at high temperature so that the charm and
anti-charm quarks inside the  charmonia indeed can be approximated by a quasi free particles. With this
idea, GR approximate the  dissociation cross section of charmonia with the
elastic cross section of charm or anti-charm quark, once the energy transfer is larger than the binding energy of charmonium.

However, as we show in the appendix, such contribution constitutes only the leading monopole contribution, whose contribution from the charm and anticharm quarks cancel as they have opposite charge.  A consistent calculation shows that the dissociation cross section is of the dipole type and only sensitive to the size of the wave function $\langle r^2 \rangle^{1/2}$ as is expected of a system composed of a quark and an antiquark system with opposite charge.

\subsection{Thermal width in the hadron gas}
The dissociation cross section of charmonium by hadron $i$ can be calculated by the following factorization formula:

\begin{equation}
\sigma_i^{diss}(s)=\sum_j \int dx~ n_{ij}(x,Q^2) ,
\label{factorization}
\end{equation}
where $n_{ij}(x,Q^2)$ is the distribution function of parton $j$ in
hadron $i$. $x$ is the longitudinal momentum fraction of parton $j$ in
hadron $i$, which is a value between 0 and 1. $Q$ is the renormalization scale of the parton distribution function. Suppose that a gluon is emitted
from a light quark in a hadron. If transverse momentum of the gluon is
smaller than $Q$, parton distribution function absorbs this
splitting process. But if the transverse momentum is larger than
$Q$, this splitting process has to be calculated in
$\sigma_j(xs,Q^2)$ of Eq. (\ref{factorization}). In other word,
dissociation cross section of charmonia by hadron $i$, the left hand
side of Eq. (\ref{factorization}), is composed of nonperturbative
part, which is the parton distribution function, and perturbative part,
which is the dissociation cross section of charmonia by parton $j$. That
is why Eq. (\ref{factorization}) is called factorization formula.

The factorization formula also shows how collinear divergence is
removed. If a massless gluon is emitted in the same or opposite direction from the original light quark or a massless gluon, then the propagator will have a pole and induce a
collinear divergence. In this case, transverse momentum of the
emitted gluon is smaller than $Q$ and thus the divergence can be absorbed by the
parton distribution function.  The elementary cross section
$\sigma_j$ as well as the parton distribution function $n_{ij}$
depends on the scale $Q^2$ but the final result, $\sigma_i^{diss}$
does not depend on the scale in principle.
There are several parton
distribution functions available. However the minimum scales $Q$ above which they are defined, are all larger
than the binding energies of charmonia, which is the renormalization scale for $\sigma_j(xs,Q^2)$.
For the case of heavy quark
scattering, the separation scale $Q$ can be taken to be  the transferred energy or the  mass of heavy quark.
But in the case of heavy quarkonium, the separation scale is the binding energy of
heavy quarkonium.

We use the Bethe-Salpeter amplitude to describe the bound state of
quarkonium in which the generated potential becomes Coulombic in the heavy quark
limit; in principle, a consistent counting and renormalization is possible in such limit.  Although the phenomenological heavy quark potential for strong force is
composed of the  Coulombic potential part and the linearly rising string potential energy \cite{Karsch:1987pv}, in the heavy quark limit, the distance between quark and antiquark is
very small and the effect of the linearly rising  string part of the potential is not felt by the quarks.  In such a limit, the binding
energy of $J/\psi$, which is 1S state, can be estimated using the mass
difference to its first excited 2S state, which will also be Coulombic.  Although the charm quark mass is not so large, we can still approximate the states to be Coulombic and obtained the binding energy of $J/\psi$ to be 780 $MeV$ \cite{Peskin:1979va,Bhanot:1979vb}.  Unfortunately, no parton distribution functions are defined in such a low $Q$.

For our purpose, we just take the  MRSSI parton distribution function
\cite{Sutton:1991ay} of the pion at $\sqrt{5}$ GeV and use it for all hadrons as the pion is the most abundant particle in the hadron gas stage.   The minimum scale $Q$ of MRSSI is still much larger than the binding
energy of $J/\psi$. In fact, due to this reason, it was found that combining the parton distribution function at larger $Q$ value than the scale of the scattering cross section in Eq.~\ref{factorization} we found inconsistent results for the charmonium states where the cross section becomes negative at some incoming energies\cite{Song:2005yd}.  Therefore, we will first apply the approach to calculate the dissociation cross section for the bottom system, perturbative QCD
approach always gives positive cross section\cite{Song:2005yd} and then extrapolate the result to the  $J/\psi$ case by the following dipole approximation,

\begin{equation}
\sigma_{J/\psi}(\sqrt{s})=\bigg(\frac{R_{J/\psi}}{R_\Upsilon}\bigg)^2\sigma_\Upsilon(\sqrt{s'}).
\label{charm-upsilon}
\end{equation}
Here, $R_{J/\psi}, R_\Upsilon$ are the radii of $J/\psi$ and $\Upsilon$ respectively.

In Coulombic potential, the radius $R= 1/\sqrt{m\epsilon_o}$, where $m$ is mass of constituent heavy quark of quarkonium, $1.94$ GeV for charm, $5.1$  GeV for bottom, and $\epsilon_o$ is the binding energy of quarkonium.  The difference in the renormalization scales of the couplings appearing in the $J/\psi$ and $\Upsilon$ cross section appearing in Eq.~(\ref{charm-upsilon}) is neglected. Two different energy scales $\sqrt{s}, \sqrt{s}'$ are assumed to have the relation $\sqrt{s'}=(m_c/m_b)\sqrt{s}$, where $m_c, m_b$ are masses of constituent charm and bottom quarks.

The upper figure of Fig. \ref{cross-width} shows the dissociation cross section of $J/\psi$ by the $\pi$ obtained using the method above.  The present cross section is still smaller than that estimated using the meson-exchange model \cite{Lin:1999ad,Oh:2000qr} or the quark-exchange model \cite{Wong:1999zb}.

The cross section obtained above  is now put to Eq. (\ref{width}) to get the thermal width of $J/\psi$ by $\pi$. For more simplicity, the thermal width in hadron gas is simplified as following:

\begin{figure}
\centerline{
\includegraphics[width=5.5cm, angle=270]{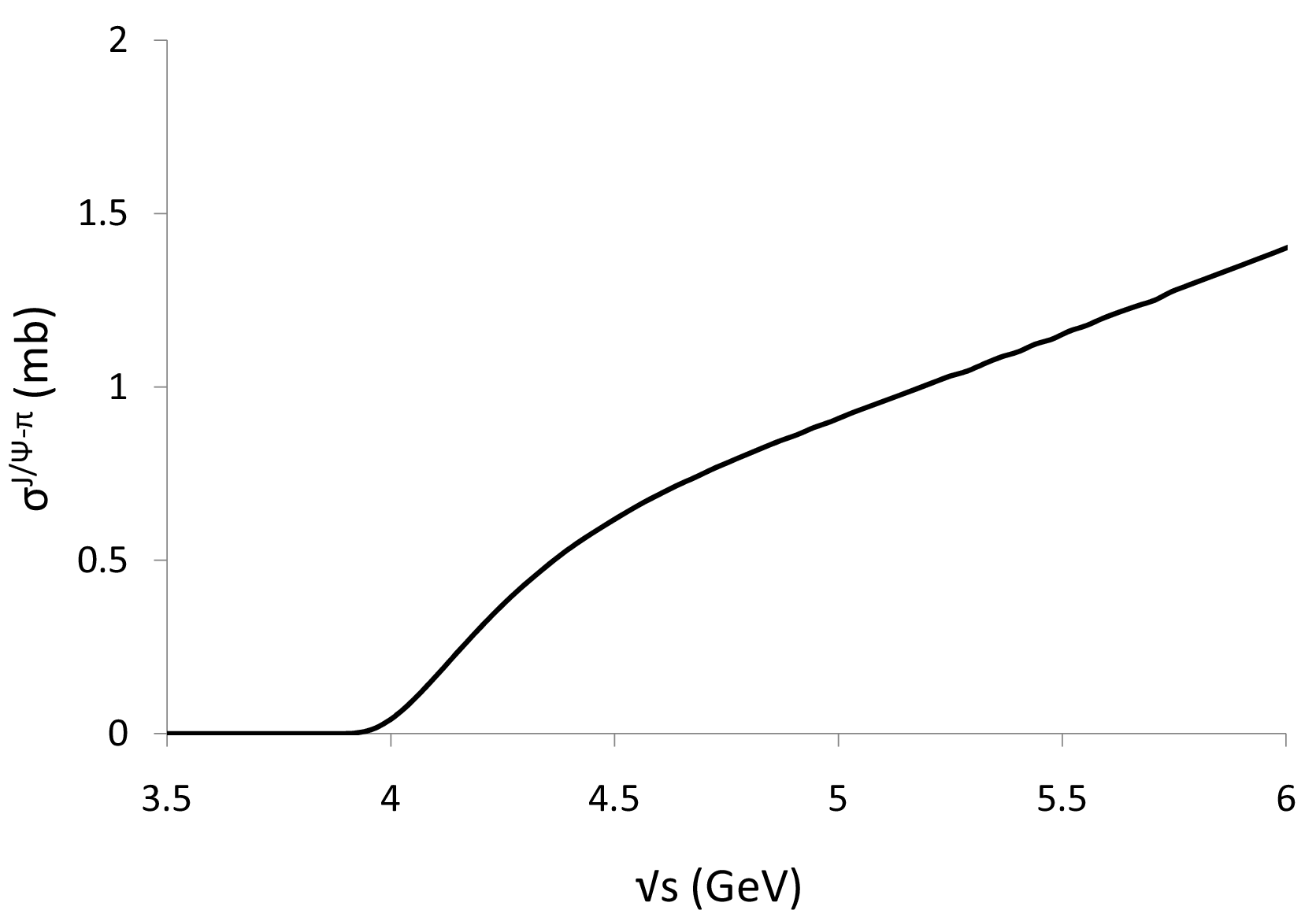}}
\centerline{
\includegraphics[width=5.5cm, angle=270]{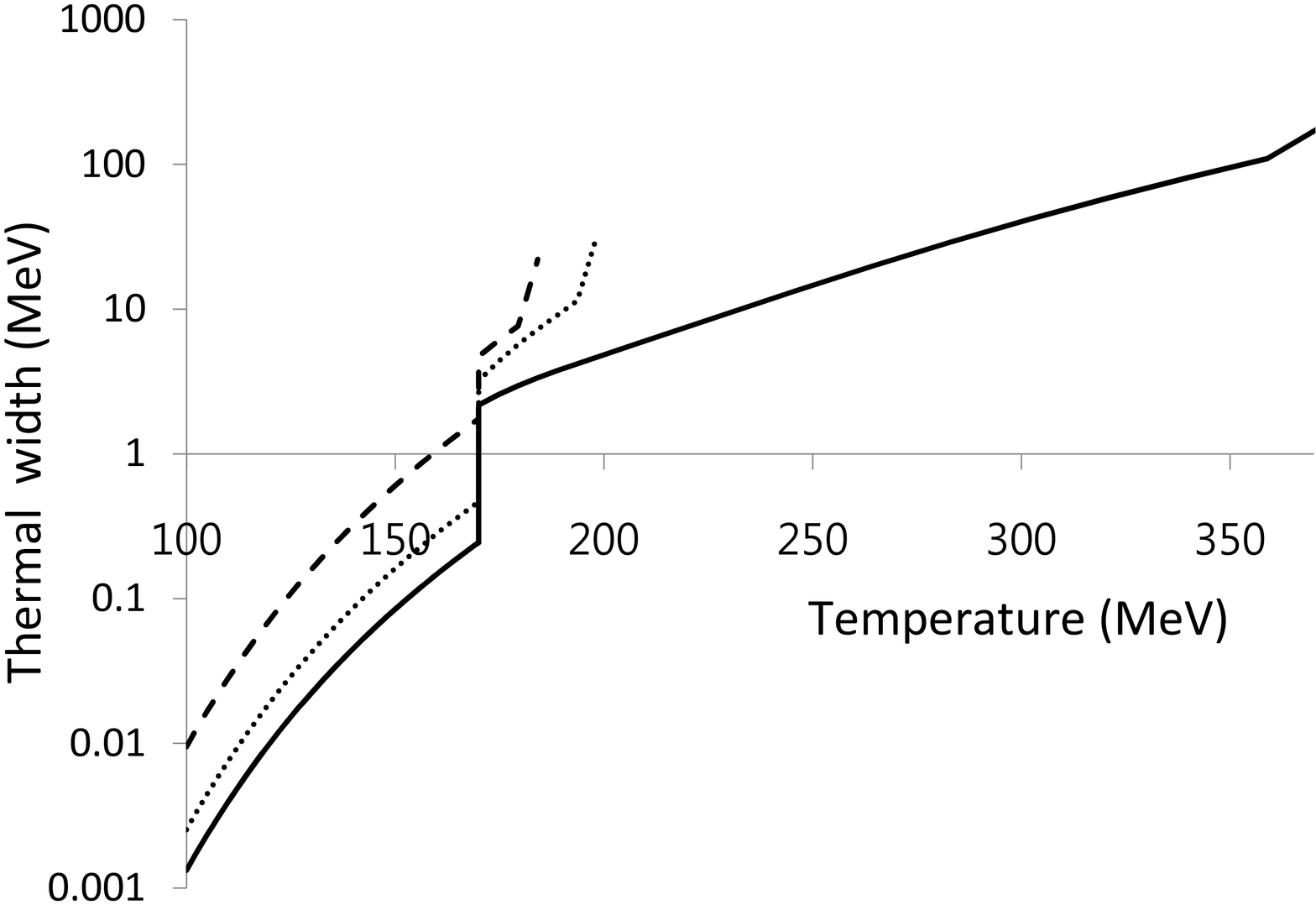}}
\caption{Upper figure: the dissociation cross section of $J/\psi$ by $\pi$. Lower figure: the thermal widths of $J/\psi$(solid line), $\chi_c$(dotted line) and $\psi'$(dashed line) in the hadron gas and in the quark-gluon plasma obtained in perturbative QCD approach.}
\label{cross-width}
\end{figure}

\begin{eqnarray}
\Gamma(T)= A \int
\frac{d^3k}{(2\pi)^3}v_{rel}(k) n_{\pi}(k,T) \sigma^{diss}(k,T),
\label{width2}
\end{eqnarray}
where $v_{rel}$ is relative velocity between charmonium and pion in rest frame of charmonium, and $\sigma^{diss}$ is dissociation cross section of charmonium by the pion. A is the ratio of number density integrated in momentum space of all hadrons, whose masses are less than 1.5 GeV for mesons and 2.0 GeV for baryons, and that of pion.
The integrated number density of pion is about $0.16\ fm^{-3}$ at temperature of chemical freeze-out, and that of all hadrons is about $0.43\ fm^{-3}$.

\begin{figure}
\centerline{
\includegraphics[width=5.5cm, angle=270]{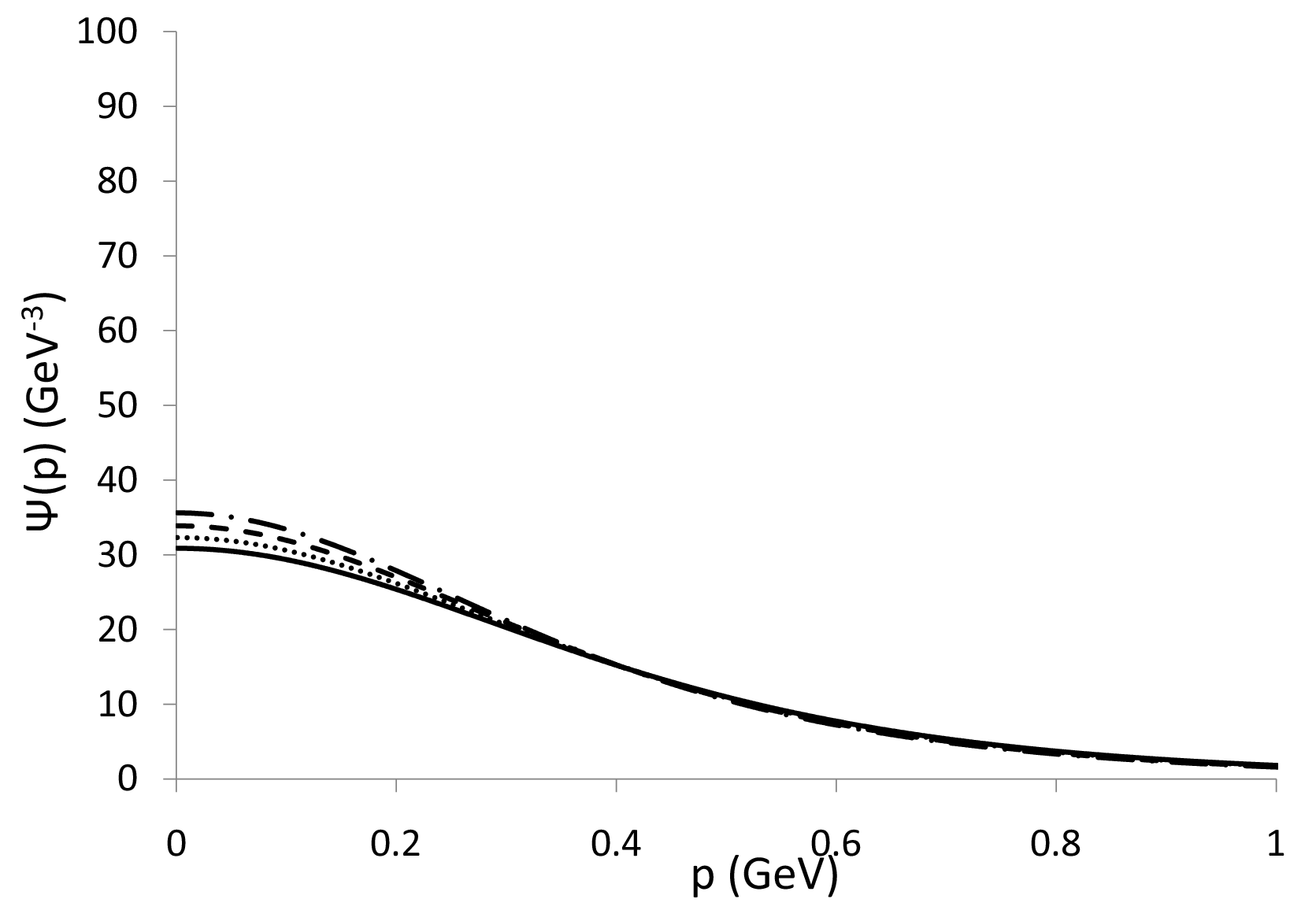}}
\centerline{
\includegraphics[width=5.5cm, angle=270]{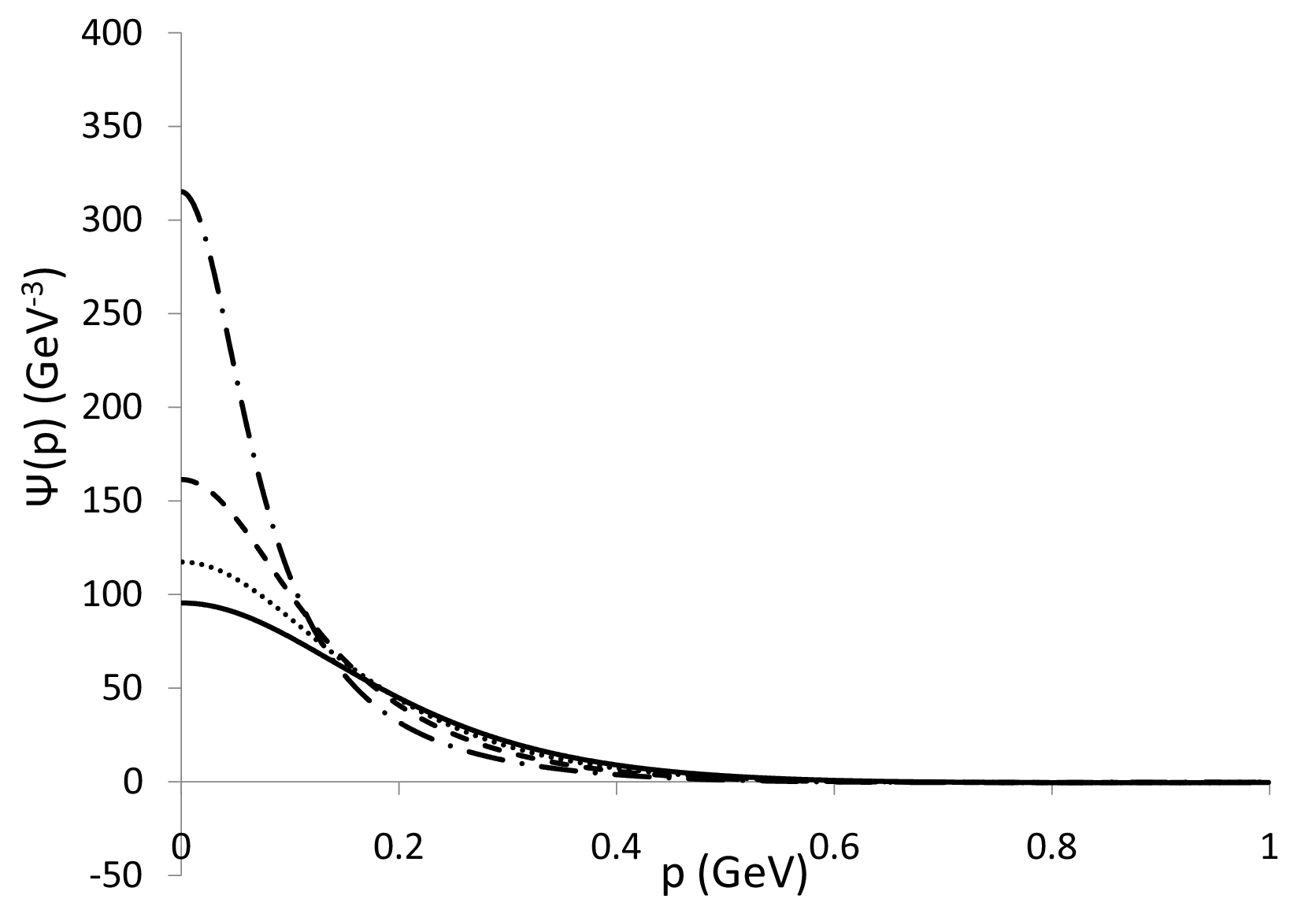}}
\centerline{
\includegraphics[width=5.5cm, angle=270]{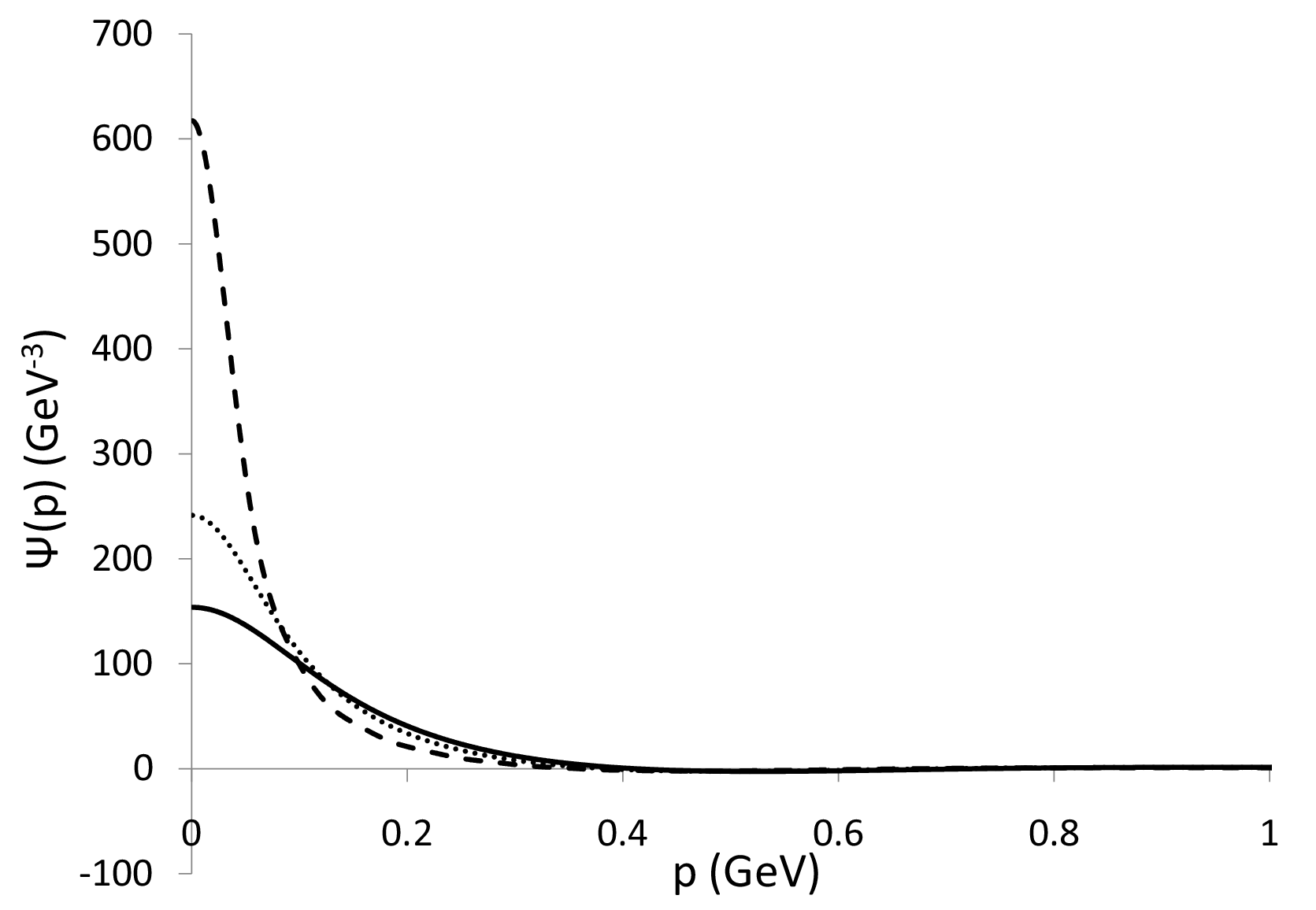}}
\caption{Wavefunctions of $J/\psi$ (upper), $\chi_c$ (middle), and $\psi'$ (lower) in momentum space, at screening masses, $289\ MeV$ (solid line), $306\ MeV$ (dotted line), and $323\ MeV$(dashed line), and $340\ MeV$ (dot-dashed line).} \label{wavefunctions}
\end{figure}

In order to calculate the dissociation cross section of charmonia in
perturbative QCD approach, we have to know their wavefunctions
\cite{Song:2005yd,Park:2007zza}. For $J/\psi$ in hadron gas,
wavefunction of 1S Coulombic bound state is used \cite{Song:2005yd}.
For excited charmonia in hadron gas, it is assumed that the cross
section is proportional to squared radius. According to the screened
Cornell potential \cite{Karsch:1987pv}, vacuum screening masses for
$J/\psi$, $\chi_c$ and $\psi'$ are $0.18,\ 0.18$, and $0.26$  GeV,
from which the radii are found to be $0.563,\ 0.778$, and $1.504\ $ fm
respectively. As a result, the dissociation cross section of
$\chi_c$ ($\psi'$) is $1.9$ (7.1) times larger than that of $J/\psi$ in the hadron gas.

\subsection{Thermal width in the QGP}

The thermal width depends on the charmonium wave function.  Here we use the Cornell potential with temperature dependent screening mass.  The form of screened Cornell potential is given as
\begin{eqnarray}
V(r,T)=\frac{\sigma}{\mu}\bigg(1-e^{-\mu r}\bigg)-\frac{\alpha}{r}e^{-\mu r},
\end{eqnarray}
where $\sigma$ is $0.192~GeV^2$ and $\alpha=0.471$.
The screening mass $\mu$ in quark-gluon
plasma depends on temperature. In the limit that $\mu$ goes to zero, the screened potential becomes the normal Cornell potential.
Fig. \ref{wavefunctions} shows the
wavefunctions of charmonia in momentum space at various screening
mass. We can see that the wavefunctions of $\chi_c$ and $\psi'$ are
sensitive to the change of screening mass, while that of $J/\psi$ is not and is more localized at the origin in momentum space.

For the temperature dependence of the screening mass, we  simply use the leading order pQCD result, $\mu=\sqrt{(N_c/3)+(N_f/6)} gT$. The coupling constant $g$ is   $1.47$ in the case of quark-gluon
plasma, which is obtained from the best fit for $R_{AA}$ data in our formalism. The details for $g$
in quark-gluon plasma will be explained later. In the case of
quark-gluon plasma, the dissociation cross section does not have
divergence, because the effective thermal masses of partons play the
role of cutoff in momentum space; in a sense the quasi particles are the  asymptotic states of the QGP  that can perturbatively scatter with the charmonium states.
As long as the temperature scale is smaller than the separation scale, such description can be thought of as an effective factorization formula, where all the soft physics is included in the quasi particles.

The lower figure of Fig. \ref{cross-width} shows the
thermal widths of charmonia both in the hadron gas and in the quark-gluon plasma, calculated using our NLO formula. We see that the
thermal widths of charmonia suddenly increase by about an order of magnitude at the
critical temperature. There are several reasons for such a sudden
increase of the thermal widths. One of them is the difference in kinetic
energies of partons.  The kinetic energy of a parton has to be larger than the binding energy of the charmonium states for the break up to take place.  Such energy are readily available for massive quarks and gluons at finite temperature.  However, for pions, the temperatures have to be higher.  Moreover, there is a large increase in the degree of freedom as phase transition occurs.

Another important factor for survival rate of $J/\psi$ is the feed-down contribution from its excited states such as the $\chi_c$, $\psi'$. Some $J/\psi$ is
produced directly but some of them is produced through the decay of
its excited states. It is not clear yet what fractions of
$J/\psi$ are produced through feed-down. Here it is assumed that
25\% and 8\% of $J/\psi$ are fed down from $\chi_c$ and $\psi'$
respectively, which is the world average \cite{LindenLevy:2009zz}. Then the
survival rate of $J/\psi$ from thermal decay is
\begin{eqnarray}
S_{th}(\vec{b},\vec{s})&=& 0.67
S^{J/\psi}_{th}(\vec{b},\vec{s})+0.25
S^{\chi_c}_{th}(\vec{b},\vec{s})\nonumber\\
&& +0.08 S^{\psi'}_{th}(\vec{b},\vec{s}), \label{decay}
\end{eqnarray}
where
\begin{eqnarray}
S^j_{th}(\vec{b},\vec{s})&=& {\rm
exp}\bigg\{-\int_{\tau_o}^{\tau_{cf}} \Gamma^j(\tau')d\tau' \bigg\}
~~~~~ (\vec{s}\in H^j) \nonumber\\
&=&0~~~~~~~~~~~~~~~~~~~~~~~~~~~~~~~~~~~~
(\vec{s}\not{\in}H^j)\nonumber.
\end{eqnarray}

The superscript $j$ in the lower equation is species of charmonia,
and $\tau_{cf}$, the upper limit of the integration, is proper time
for the chemical freeze-out. $H^j$ is the hot region where initial
temperature is over the dissociation temperature of each charmonia.

\section{the regeneration of charmonia}
As discussed before, results  from hydrodynamics suggest that hot nuclear matter created in relativistic heavy ion collision is thermalized very fast.
Moreover particle yields and their ratios from relativistic heavy ion collision are well reproduced using the grand canonical ensemble.
If the number of charm  and anti-charm quarks are also thermalized, the total number of charm pairs produced $N_{c\bar{c}}^{AB}(\tau)$ from the collision of nucleus A and B would satisfy the following equation \cite{BraunMunzinger:2000px}:

\begin{eqnarray}
N_{c\bar{c}}^{AB}(\tau)=\bigg\{\frac{1}{2}n_o(\tau)+n_h(\tau)\bigg\}V(\tau).
\label{recom1}
\end{eqnarray}
Here, $n_o$ and  $n_h$ are the number density of open and  hidden charms respectively, and $V$ is the volume of hot nuclear matter; all are a function of the proper time $\tau$.

In general, if creation or annihilation of charm pairs is considered during the expansion of hot nuclear matter, the number of charm quark pairs will be a function of proper time $\tau$.  Moreover, if the lifetime of hot nuclear matter is sufficiently longer than the thermalization time,  the number of charm pairs will become thermalized and satisfy Eq. (\ref{recom1}); thermalization will proceed  through the processes such as $c+\bar{c}\leftrightarrow q+\bar{q},\ c+\bar{c}\leftrightarrow g+g$. However, estimates based on perturbative QCD suggests that the lifetime of hot nuclear matter is not long enough for the number of charm pairs to be thermlized, because the cross section for creation or annihilation of charm pairs is small with respect to the life time of the fireball.

\begin{figure}
\centerline{
\includegraphics[width=5.5cm, angle=270]{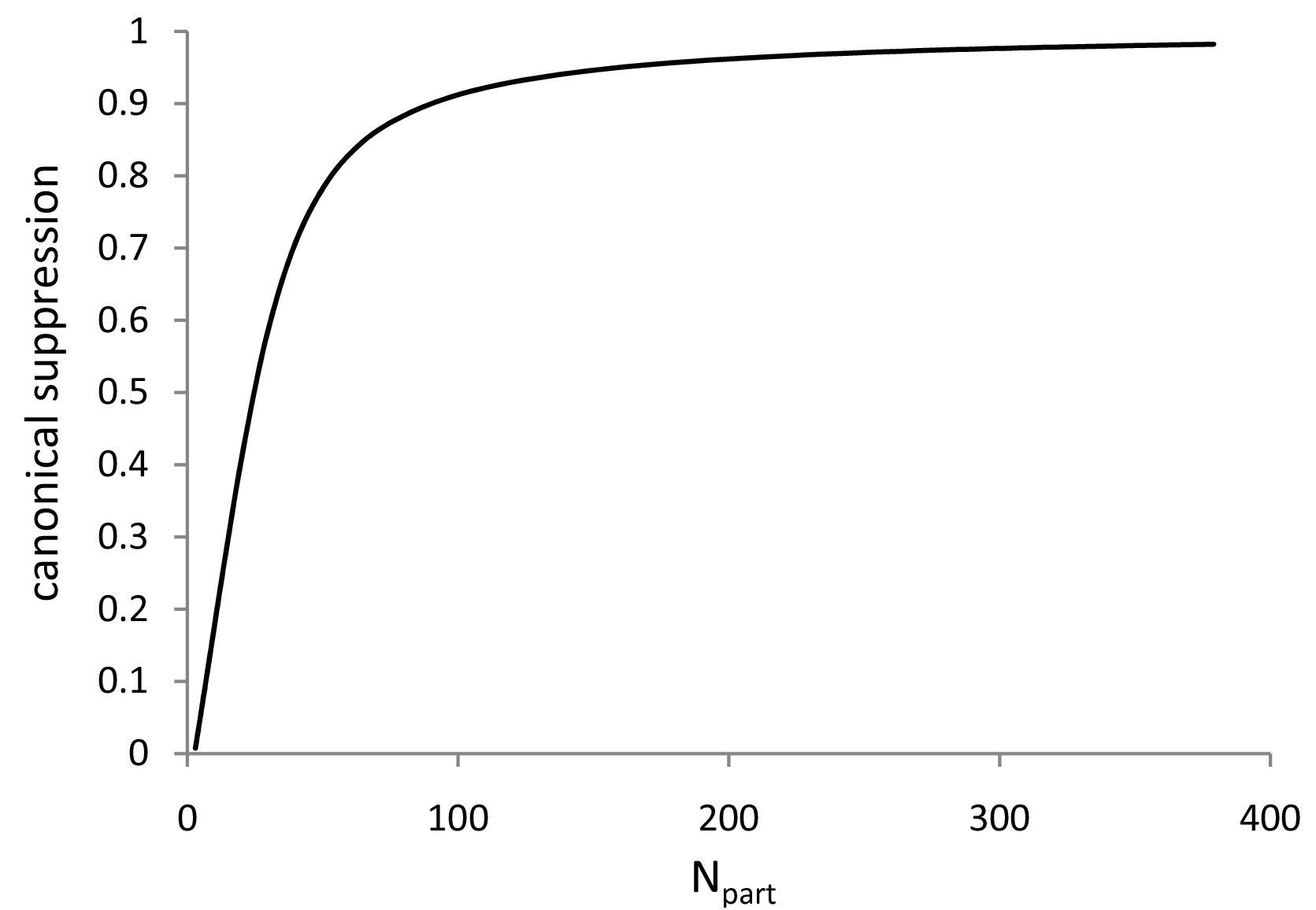}}
\centerline{
\includegraphics[width=5.5cm, angle=270]{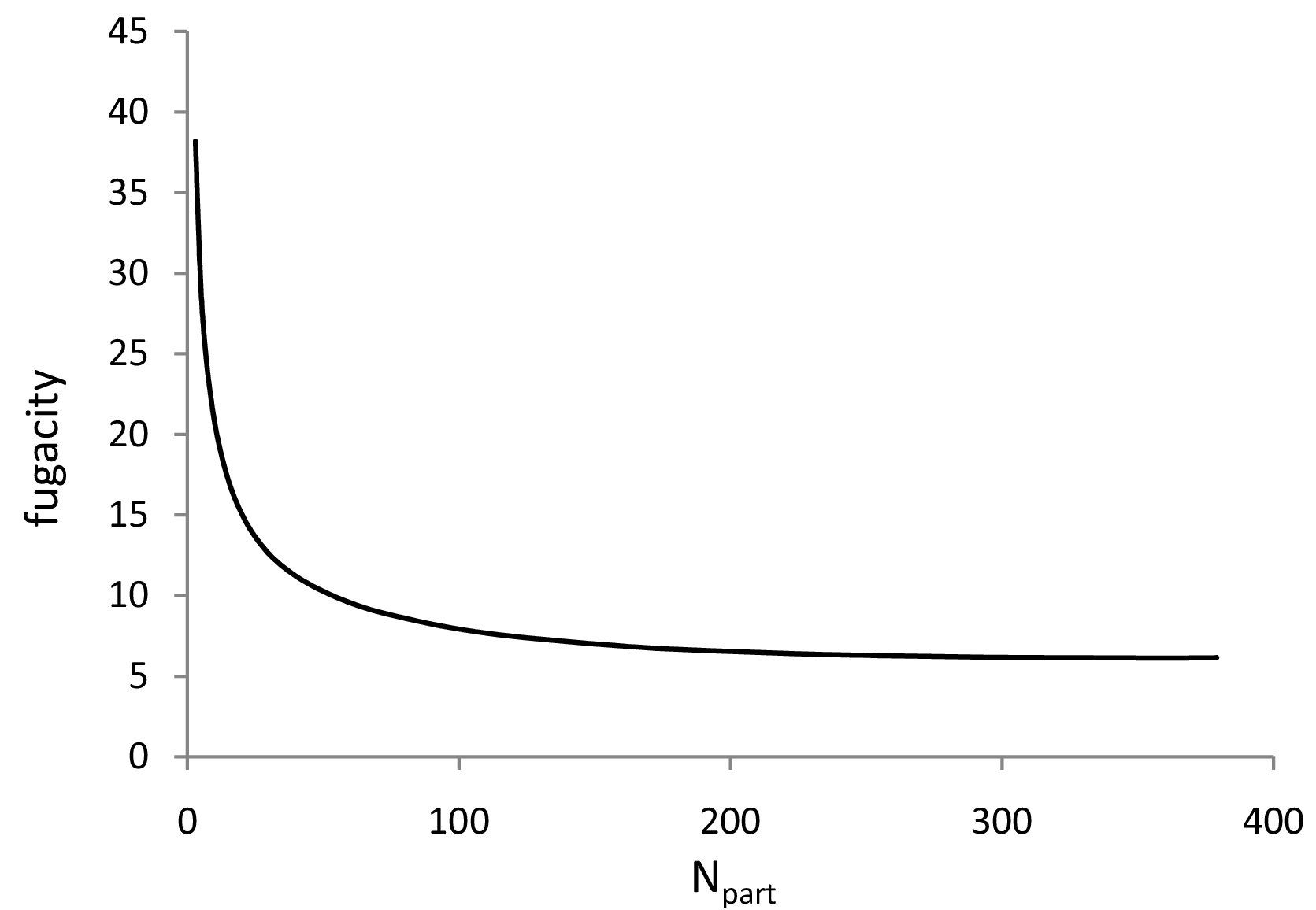}}
\caption{Canonical suppression factor(upper) and fugacity(lower) with respect to number of participants}
\label{canonical}
\end{figure}

Therefore, taking into account the off-equilibrium effects for the heavy quarks, the number of charm pairs given in  Eq. (\ref{recom1}) will be  modified as follows,

\begin{eqnarray}
N_{c\bar{c}}^{AB}(\tau)=\bigg\{\frac{1}{2}\gamma n_o(\tau)+\gamma^2 n_h(\tau)\bigg\}V(\tau),
\label{recom2}
\end{eqnarray}
where $\gamma$ is called the fugacity \cite{BraunMunzinger:2000px}. If
the number of charms is more (less) then what it would be if the
charm quarks are  thermalized, fugacity is larger (smaller) than
unity. In the central collision of  RHIC the initial temperature of
created hot nuclear matter is very high and the  fugacity at
$\tau=\tau_0$ is smaller than 1.  As the system cools down, the
fugacity increases.  This fact can in fact be simply understood
qualitatively as follows; assuming isentropic expansion, $V(\tau)
\propto 1/n_L$ where $n_L$ is the density of light partons,
substituting this into Eq.~(\ref{recom2}) gives
$N_{c\bar{c}}^{AB}(\tau) \propto \bigg\{\frac{1}{2}\gamma
\frac{n_o(\tau)}{n_L}+\gamma^2 \frac{n_h(\tau)}{n_L} \bigg\}$, since
$\frac{n_o(\tau)}{n_L} \approx exp(-m_c/T)$, where $m_c$ is the mass
of charm quark, and assuming the number of charm quarks do not
change during the expansion, we can conclude that the fugacity
should increase as the temperature decreases during the expansion.
The final value of fugacity at $T_c$ as a function of $N_{\rm part}$
is shown in the lower figure of Fig. \ref{canonical}.

In Eq.
(\ref{recom2}) the number density of open charm is multiplied by
$\gamma$ and hidden charm by $\gamma^2$, because open charm has one
charm or anti-charm quark while hidden charm has both charm and
anti-charm quarks. It may be assumed that the number of charm pairs
does not change during the expansion of hot nuclear matter, because
the creation and annihilation cross section of the pair is very
small. Then the left hand side of Eq. (\ref{recom2}) dose not change
and its initial value is calculable with Glauber model as following:
\begin{eqnarray}
N_{c\bar{c}}^{AB}(\vec{b})&=&\sigma_{c\bar{c}}^{NN} AB \int d^2 s T_A(\vec{s}) T_B(\vec{b}-\vec{s})\nonumber\\
&=&\sigma_{c\bar{c}}^{NN} AB T_{AB}(\vec{b}),
\label{ncc}
\end{eqnarray}
where $\sigma_{c\bar{c}}^{NN}$ is the cross section for $c\bar{c}$ pair production in p+p collision, which is $63.7\ \mu b$ per rapidity near midrapidity at $\sqrt{s}=200$ GeV in perturbative QCD \cite{Cacciari:2005rk,Andronic:2006ky}.

Another correction is about system size. So far grandcanonical
ensemble is used for number densities of particles. Canonical
ensemble, however, is to be used technically, because quantum
numbers of the system are to be conserved. Quantum numbers of the
system are fixed in canonical ensemble, while they are not in
grandcanonical ensemble. The function of probability density with
respect to quantum number is a delta function in canonical ensemble,
while it is a  gaussian type distribution in the grandcanonical ensemble.
As the system size is large, the width of function of probability
density in grandcanonical ensemble is narrow. In the thermodynamical
limit, where the system size is infinite, the function becomes like
a delta function. Therefore, grandcanonical ensemble and canonical
ensemble are equal in thermodynamical limit. The equality seems to
be true in central collisions of heavy ions, because the number of
produced charm or anticharm are not small. However it is questionable for peripheral collisions. The quantum number in grandcanonical
ensemble can be converted to that in canonical ensemble by simply
multiplying modification factors as follows;

\begin{eqnarray}
N_{c\bar{c}}^{AB}&=&\bigg\{\frac{1}{2}\gamma n_o(\tau)\frac{I_1(\gamma n_o(\tau) V(\tau))}{I_o(\gamma n_o(\tau) V(\tau))}\nonumber\\
&&~~~~~~~~~~~~~~+\gamma^2 n_h(\tau) \bigg\}V(\tau),
\label{fugacity}
\end{eqnarray}
where $I_o$ and $I_1$ are modified Bessel functions and their ratio that depends on the number of charm quarks is called the canonical ensemble correction \cite{Gorenstein:2000ck}.

The upper figure of Fig. \ref{canonical} is the canonical ensemble corrections at hadronization. In central collisions, the number of produced charm quarks are not small so that the correction is almost unity. However the correction rapidly decreases in peripheral collisions because the number of produced charm quarks becomes small. In other words, the difference between grandcanonical ensemble and canonical ensemble becomes serious.

Fugacity $\gamma$ is calculated from  Eq. (\ref{ncc}) and (\ref{fugacity}). The lower figure of Fig. \ref{canonical} shows the fugacity at hadronization. The Fugacity is about $6$ at central collision and remains constant until $N_{part}$ decreases to about 100, and then rapidly increases.
The rapid decrease of canonical ensemble correction causes  fugacity to rapidly increase at small $N_{part}$.

So far it is assumed that charm quarks are thermalized kinetically. In other words, it is assumed that the kinetic energy of the charm quarks follow thermal distribution. This thermalization is attained through elastic scattering of charm quarks with the surrounding partons. However the elastic, as well as the inelastic, cross sections of the heavy quarks are smaller than those of the light quarks.   The formation of charmonium states through recombination are more probable when the charm and anti-charm quarks are randomly distributed in momentum space.   Therefore, to estimate the regeneration effect, it is important to know how much charm quarks and anti-charm quarks are thermalized kinetically at the hadronization point.

To estimate this effect, one defines the relaxation factor as follows, \cite{Grandchamp:2002wp}:

\begin{equation}
R\equiv 1-{\rm exp}\bigg(
-\int_{\tau_o}^{\tau_{QGP}}\frac{d\tau}{\tau_{eq}}\bigg),
\label{relaxation}
\end{equation}
where $\tau_{QGP}$ is the proper time the quark-gluon plasma phase terminates. Relaxation time, $\tau_{eq}$, is defined as follows:

\begin{equation}
\tau_{eq}(T)\equiv 1/\bigg( \sum_i \int \frac{d k^3}{(2\pi)^3} v_{rel}(k) n_i(k,T) \sigma_i(k)\bigg),
\end{equation}
where $n_i$ is the number density of parton $i$ in the quark-gluon
plasma, and $\sigma_i$ is the elastic cross section of charm or
anti-charm quark by parton $i$ in the charm or anti-charm rest frame.
The cross section multiplied by number density of partons is the
mean free path of charm or anti-charm quark. Therefore $\tau_{eq}$
is the average elastic collision time of charm or anti-charm quarks
with the light partons. The elastic cross section is calculated to
leading order in perturbative QCD. The effective thermal masses
of partons extracted from lattice by Levi and Heinz are used
\cite{Combridge:1978kx,Levai:1997yx}. The relaxation factor $R$ is
about $0.25$ at central collision and decreases as the collision
becomes peripheral.

The number of regenerated $J/\psi$ can now be written as

\begin{eqnarray}
N_{reg~J/\psi}^{AB}&=&\gamma^2 \bigg\{n_{J/\psi}S_{th-H}^{J/\psi} +Br(\chi_c\rightarrow J/\psi)n_{\chi_c}S_{th-H}^{\chi_c}\nonumber\\
&&~~~~+Br(\psi'\rightarrow J/\psi)n_{\psi'}S_{th-H}^{\psi'}\bigg\}VR,
\end{eqnarray}
where $Br(\chi_c\rightarrow J/\psi)$ and $Br(\psi'\rightarrow J/\psi)$ are branching ratios from $\chi_c$ to $J/\psi$ and from $\psi'$ to $J/\psi$ respectively.
Canonical ensemble correction is not multiplied, because charmonia are hidden charms.
$S_{th-H}^{J/\psi}$, $S_{th-H}^{\chi_c}$ and $S_{th-H}^{\psi'}$ are respectively the survival rates of $J/\psi$, $\chi_c$ and $\psi'$ from thermal decay in hadron gas defined as follows:

\begin{eqnarray}
S_{th-H}^i={\rm exp}\bigg\{-\int_{\tau_H}^{\tau_{cf}} \Gamma^i(\tau')d\tau' \bigg\}.
\end{eqnarray}
Here, $\tau_H$ is the proper time the hadronization is completed.

\section{results}
$J/\psi$ is a massive particle produced through hard collision of two nucleons from each colliding nucleus. If there is no nuclear modification, the number of produced $J/\psi$ in a heavy ion collision will be proportional to the number of binary collisions  as follows,

\begin{equation}
N_{J/\psi}^{\rm Glauber}(\vec{b})=\sigma_{J/\psi}^{NN}A^2 T_{AA}(\vec{b}).
\label{no-modification}
\end{equation}
Here, $\sigma_{J/\psi}^{NN}$ is the cross section for $J/\psi$ production in p+p collision, which is recently measured to be $0.774\ \mu b$ per rapidity near midrapidity at $\sqrt{s}=200$ GeV \cite{Adare:2006kf,Andronic:2006ky}. $T_{AA}$ is the  thickness function defined in Eq. (\ref{two-scales}).

The nuclear modification factor $R_{AA}$ is defined as follows,
\begin{eqnarray}
R_{AA}(\vec{b})=\frac{N_{J/\psi}^{AA}(\vec{b})}{\sigma_{J/\psi}^{NN} A^2 T_{AA}(\vec{b})}.
\label{recom}
\end{eqnarray}

The numerator in the right hand side of the above equation is the actual number of
$J/\psi$ produced in A+A collision, while the denominator is the
expected number of $J/\psi$ production in the same collision. However,
nuclear absorption and thermal decay of charmonia suppress the
actual number of produced $J/\psi$ so that these effects supress
$R_{AA}$. On the contrary, regeneration of charmonia enhances the
number so that it raises $R_{AA}$. Considering these effects
altogether, the nuclear modification factor is from Eq.
(\ref{absorption}), (\ref{decay}) and (\ref{recom})
\begin{eqnarray}
R_{AA}(\vec{b})= \int d^2 s S_{nuc}(\vec{b},\vec{s})
S_{th}(\vec{b},\vec{s}) +\frac{N_{reg
~J/\psi}^{AA}(\vec{b})}{\sigma_{J/\psi}^{NN}A^2 T_{AA}(\vec{b})}.
\label{raa}
\end{eqnarray}
The first term of the right hand side represents nuclear absorption and thermal decay of $J/\psi$, which includes the convolution in the transverse plane that takes into account the initial suppression if the local temperature is larger than the charmonium dissociation temperature, and the second term regeneration of $J/\psi$.

\begin{figure}
\centerline{
\includegraphics[width=5.5cm, angle=270]{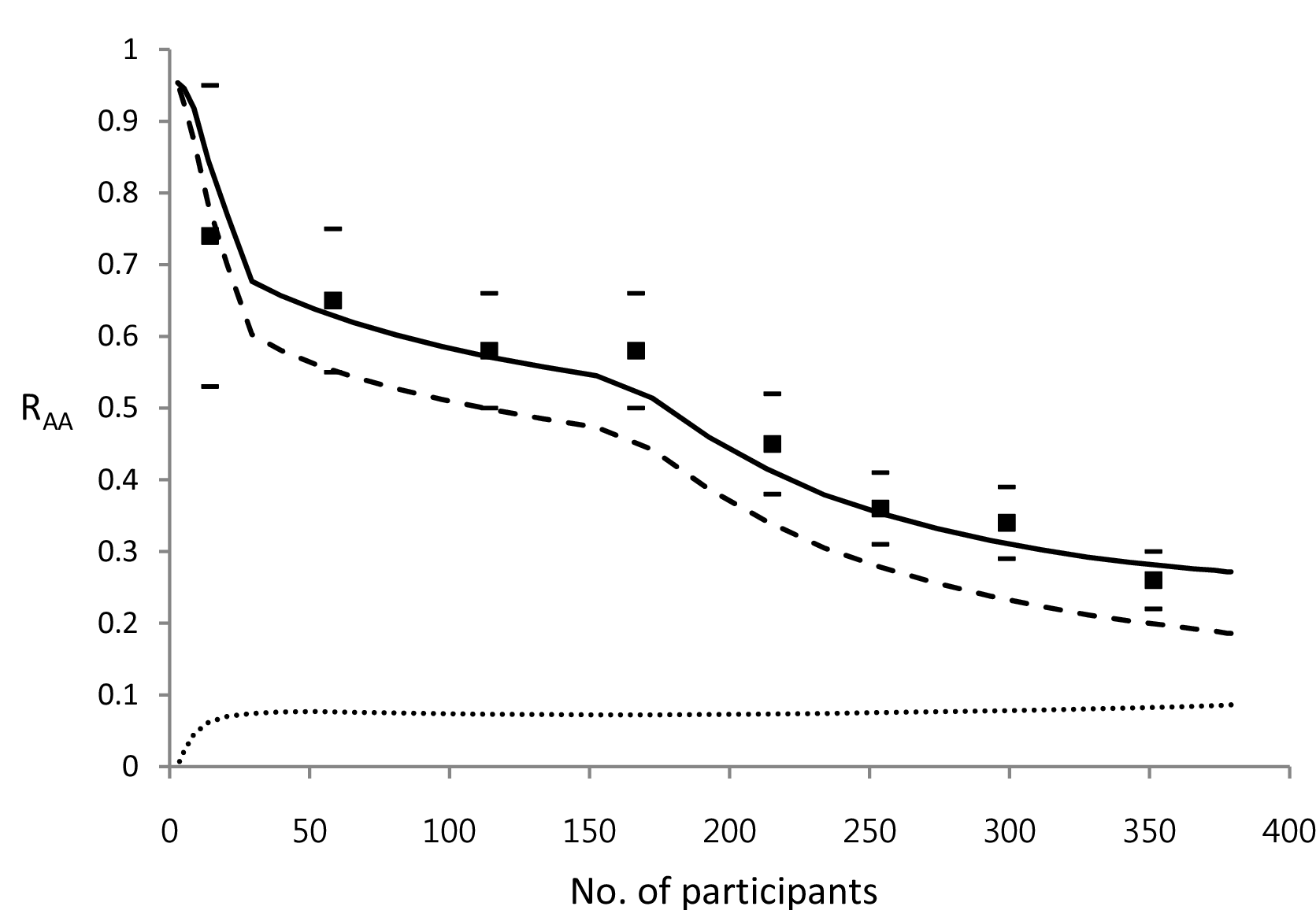}}
\centerline{
\includegraphics[width=5.5cm, angle=270]{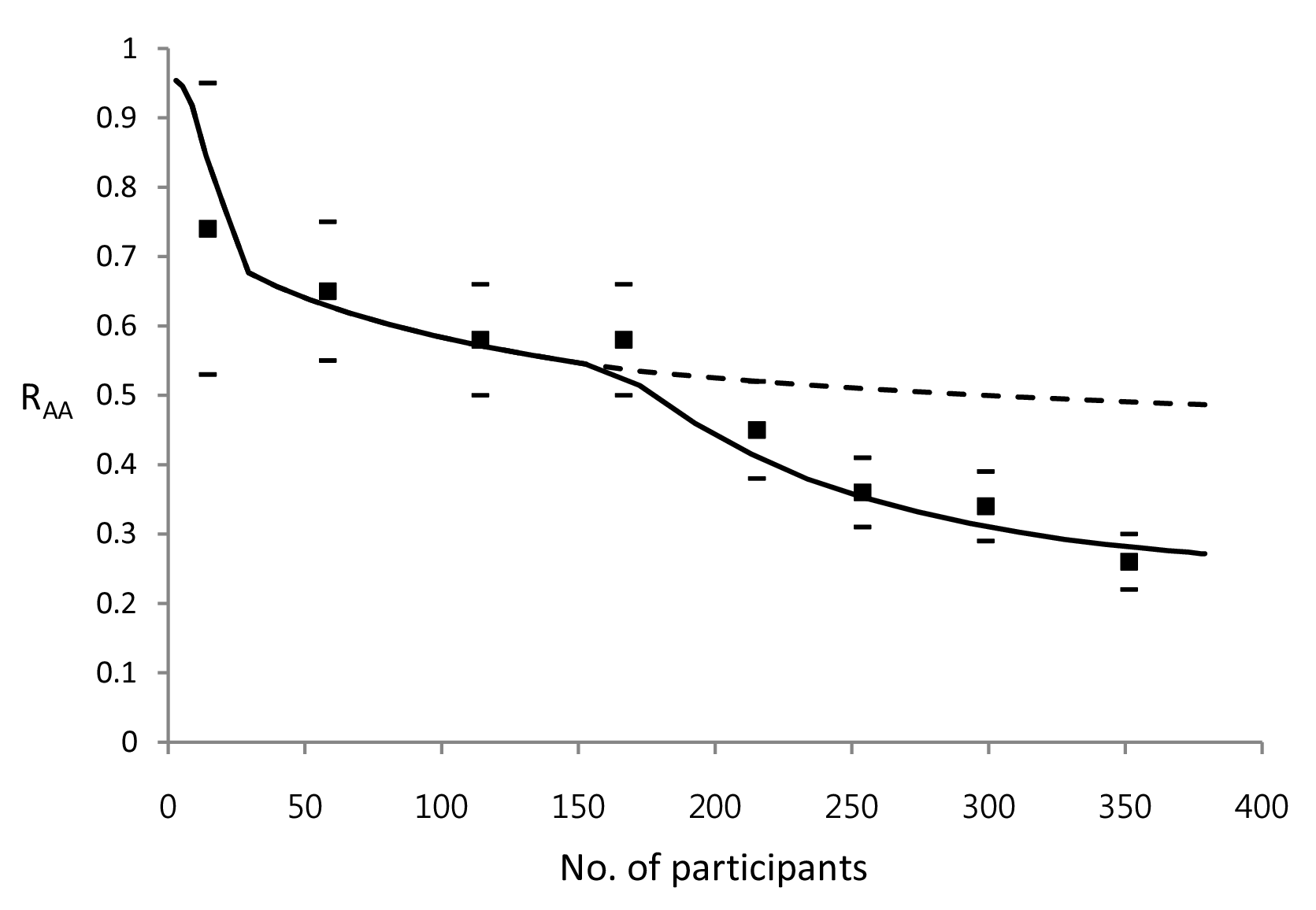}}
\caption{The upper figure compares $R_{AA}$ from the experiment
\cite{Adare:2006ns} to that calculated from Eq. (\ref{raa}). The dashed
line is the  nuclear absorption and thermal decay, the
dotted line the  regeneration effect and the solid line
their sum. As for the lower figure, the experimental data and the solid line is the same as in the upper figure, but the dashed line shows the result when the initial suppression of $J/\psi$ in the hot region is not taken into account.} \label{raaf}
\end{figure}

\subsection{results for Au+Au}

The upper figure of Fig. \ref{raaf} shows that Eq. (\ref{raa}), with a best fit coupling $g=1.47$, reproduces
the experimental data of $R_{AA}$ \cite{Adare:2006ns} well. The
dashed line is $S_{nuc}\ S_{th}$, the dotted line $R_{AA}^{reg}$ and
the solid line their sum.  The coupling constant $g$ is an
important parameter in the result. It determines the thermal widths
of charmonia, their dissociation temperatures and the
relaxation factor $R$. When the coupling constant increases, the
thermal widths of charmonia become wider, the dissociation
temperatures of charmonia become lower and the relaxation factor $R$
larger, and vice versa.  While perturbative QCD formulas are used, it is known that the quark-gluon plasma
from $T_c$ up to about $2\ T_c$ is nonperturbative. In this sense,
the coupling constant $g$ is kind of a free parameter fitted to
reproduce the experimental data.

In our approach the value of the
coupling $g$ also determines the position of a kink on $R_{AA}$
curve.  At present, the experimental uncertainties in $R_{AA}$ is
still too large to extract any detailed structure as a function in
$N_{\rm part}$.  Nevertheless, taking the central experimental data
of $R_{AA}$ seriously, there seems to be two drops. The first one
appears at small number of participants, and the second one between
$170$ and $210$ $N_{\rm part}$.

There are two important thermal
effect on charmonia production: one is the initial suppression which
occurs when the local initial temperature of the hot region is
higher than the dissociation temperature, the other is thermal
decay of charmonia through interaction with surrounding particles.
Two effects occur at different temperatures, which are well
distinguished in the lower figure of Fig. \ref{cross-width}.   From
the critical temperature to the dissociation temperature, which are
respectively identified by the sudden increase and the divergence
of the thermal width, the thermal width of charmonium increases
monotonically; in the lower figure of Fig. \ref{cross-width}, the
divergence is represented by the termination point. For the initial
suppression effect, the survival rate of charmonium is 0, while for
the effect coming from thermal decay, it depends on the temperature and
lifetime of the QGP relevant for thermal decay.

It should be noted that
the survival rate will change appreciably only when the thermal
width becomes large, because the lifetime of quark-gluon plasma in
heavy ion collision is not so long. In fact, as can be seen from Fig. \ref{temp-potentials}, the temperature region near $T_c$ is more important, which last almost 2 to 3 fm/c.  This means that the thermal width has to be of the order 100 MeV to have an appreciable suppression effect.  Therefore, there will be a
critical change in the survival rate depending on whether or not the
initial temperature is larger than the dissociation temperature;
because that condition corresponds to having either an infinite or
finite thermal width. This critical change in survival rate will
appear as a sudden change of $R_{AA}$ along the axis of number of
participants, because the initial temperature rises as the number of
participants increases. Therefore, if at some future point, the error bars in the
experimental data are reduced and one indeed finds the
sudden kink, it would strongly imply that the $J/\psi$ dissociation
takes place at the corresponding point. Moreover, an initial kink
could appear when the initial temperature is higher than the
dissociation temperatures of excited charmonia, such as $\chi_c$ and
$\psi'$.  It should be kept in mind however that when the thermal decay width is in the order of few hundred MeV just above the deconfinement, the sudden kink could also have occurred due to the thermal decay.

Again, taking the central values of $R_{AA}$ seriously,  the second kink can be translated to a dissociation temperature of $386~MeV$;
this can be obtained from  Fig. \ref{profile} and the Glauber model
given in Eq.~(\ref{two-scales}) that links $N_{\rm part}$ to the
impact parameter.  Now, solving Schrodinger equation with the screened Cornell potential, one finds that the dissociation occurs when the screening mass is  about $\mu=695~MeV$.  Since the  screening mass is simply $\mu=\sqrt{(N_c/3)+(N_f/6)}gT$ in our calculation, the
coupling constant $g$ has to be fixed at $1.47$ to reproduce the screening mass at the dissociation temperature $386~MeV$.  This coupling relates to $\alpha=0.172$ which is not so different from the lattice calculation for the asymptotic value of the coupling reached at higher temperature\cite{Kaczmarek} and to the coupling given in Eq.~(\ref{lhcoupling}).

With the same value of $g$, one finds the dissociation temperatures of $\chi_c$ and $\psi'$ to be $199~MeV$
and $185~MeV$ respectively.  The initial drop of the solid line in both the upper and lower figures of Fig.
\ref{raaf} is coming from the initial suppression of both $\chi_c$ and $\psi'$.
The dashed line of the lower figure of Fig.
\ref{raaf} corresponds to the case where the initial suppression of $J/\psi$  is
not taken into account. We can see how much $R_{AA}$ is improved by
considering the effects of the  initial hot region.

\subsection{understanding the results}

\begin{figure}
\centerline{
\includegraphics[width=6 cm, angle=270]{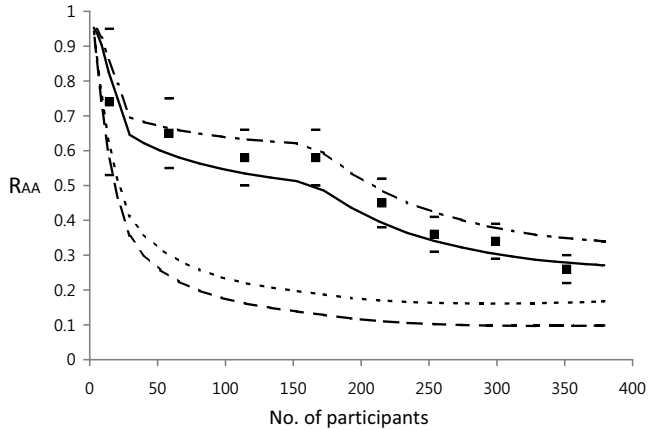}}
\caption{$R_{AA}$ when the thermal width and mass are artificially changed.  The different curves are explained in the text.  }
\label{RHIC_Au}
\end{figure}

To gain some more insights into what causes the characteristic features of the solid line in Fig. \ref{raaf}.   Let us artificially change some parameters and investigate the changes in the prediction.  The results are shown in Fig. \ref{RHIC_Au}. The solid line is the best fit value using the same method as described above but obtained with $g=1.47$  and with a constant thermal width of 10 MeV.  The dashed and dot-dashed curves are when the thermal width is changed to 100 MeV and zero respectively.  One sees that the slope of the curve critically depends on the thermal width. While the increase in the width suppresses $R_{AA}$, one should also note that an increase in the width also implies a larger coupling to the medium and hence an increase in the relaxation factor.  This will enhance the regeneration effects and thus compensate the suppression.

In the dotted curve, the mass of $J/\psi$ is artificially changed by -100 MeV. This is to see if the sudden mass shift of $J/\psi$ recently advocated by Morita and one of us\cite{Morita:2007pt} will have any effect on $R_{AA}$.  As one can see, due to the enhancement of the statistical factor in regeneration effect, the curve is pushed up by a nontrivial amount.  Therefore,  the effects of a sudden decrease in the mass and a sudden increase in the width at the phase transition point, as advocated in Ref.~ \cite{Lee:2008xp,Song:2008bd},  might just compensate each other in $R_{AA}$, making it difficult to identify such effects from a measurement of $R_{AA}$ alone.

\begin{figure}
\centerline{
\includegraphics[width=5.5cm, angle=270]{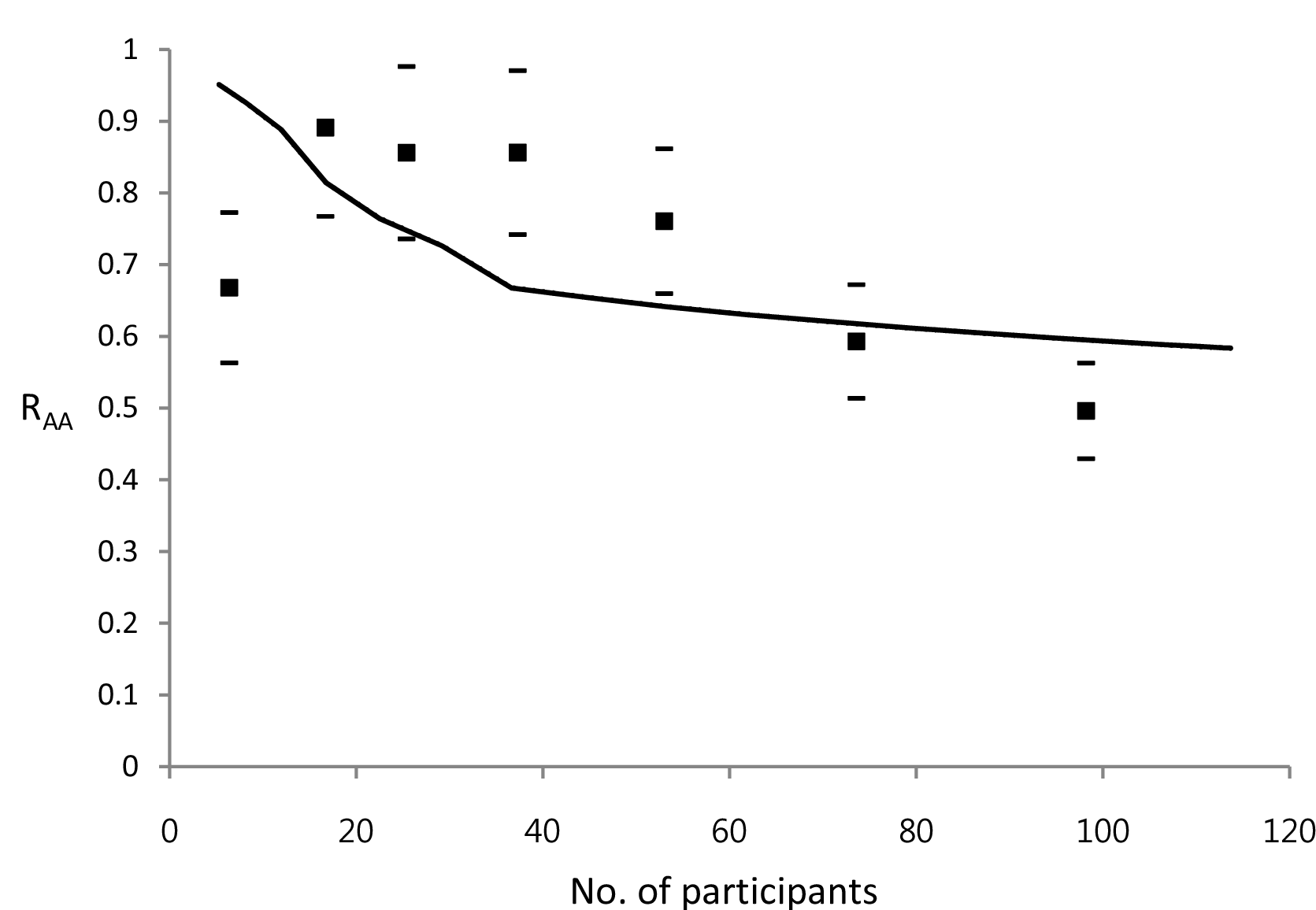}}
\centerline{
\includegraphics[width=5.5cm, angle=270]{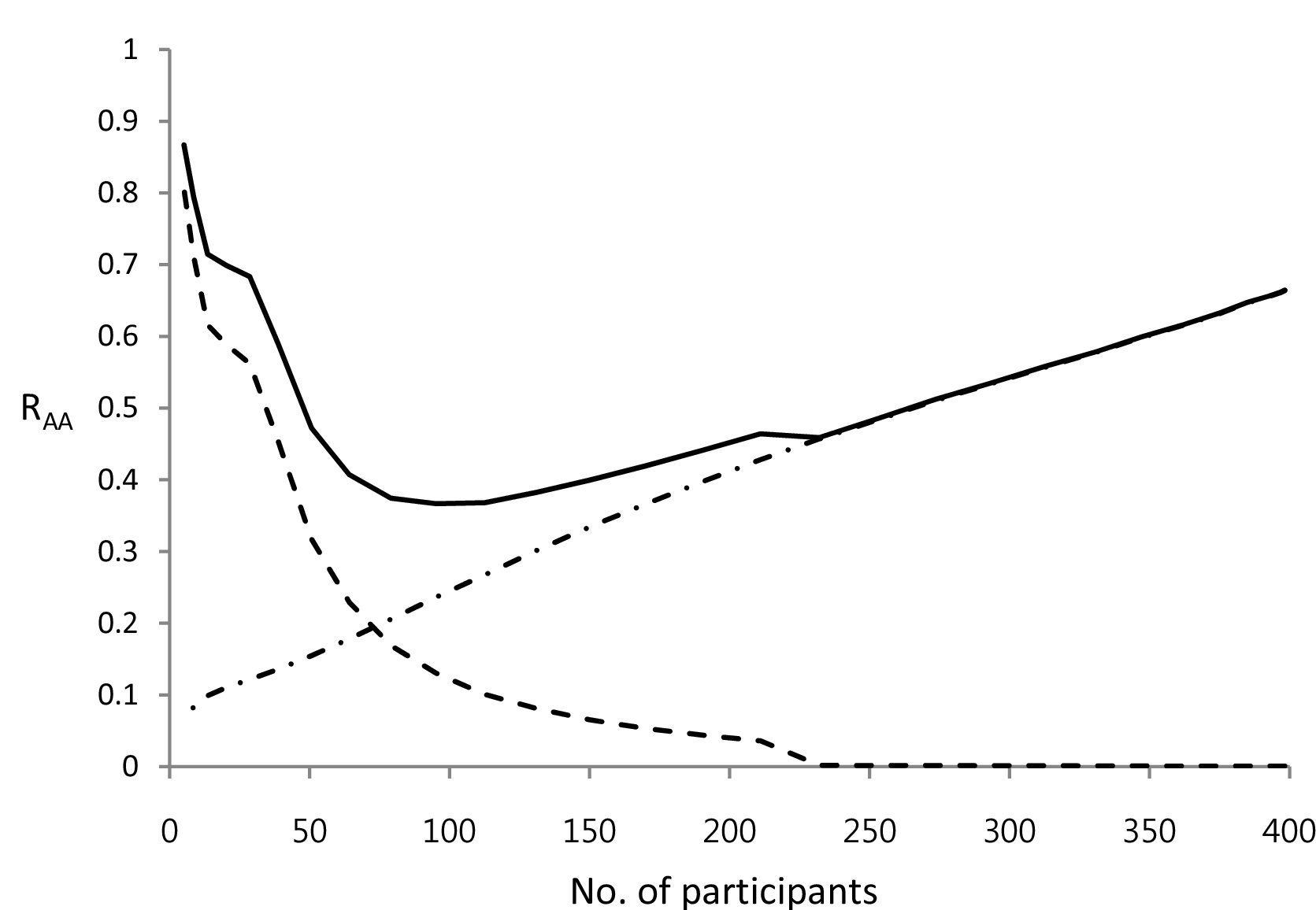}}
\caption{The upper figure is $R_{AA}$ of $J/\psi$ in Cu+Cu collision at $\sqrt{s_{NN}}=200$ GeV. The solid line for $a_{\bot}=0.1~c^2/$ fm and the dotted line for $a_{\bot}=0.08~c^2/$ fm are overlapped. The lower figure is $R_{AA}$ of $J/\psi$ in Pb+Pb collision at $\sqrt{s_{NN}}=5.5$ TeV. The dashed line is $R_{AA}$ after nuclear absorption and thermal decay, the dot-dashed line is $R_{AA}$ from regeneration effect and the solid line is their sum.}
\label{others}
\end{figure}

\subsection{results for Cu+Cu}

The same method is applied to Cu+Cu collision at $\sqrt{S_{NN}}=200$ GeV. The only difference from the previous analysis is the geometry of the colliding nuclei. $r_o$ and $C$ for Cu are set at $4.163~$ fm and $0.606~$ fm respectively in Eq. (\ref{Woods-Saxon}) \cite{De Jager:1987qc}. The upper figure of Fig. \ref{others} shows $R_{AA}$ of $J/\psi$ in Cu+Cu collision at $\sqrt{S_{NN}}=200$ GeV. Transverse acceleration of hot nuclear matter in Cu+Cu collision will be less than that in Au+Au collision, so that two different transverse accelerations, $0.1~c^2/$fm and $0.08~c^2/$fm, are used in the calculation for $R_{AA}$. However there is almost no difference in $R_{AA}$ so that the two curves almost overlap in  Fig. \ref{others}. The calculation for the Cu+Cu collision slightly overestimates the experimental data at central collisions while it underestimates at peripheral collisions. In Cu+Cu collision at $\sqrt{S_{NN}}=200$ GeV, the maximum temperature is lower than the dissociation temperature of $J/\psi$ even at the most central collision, hence no kink appears in $R_{AA}$.

\subsection{results for LHC}

We can attempt to  predict $R_{AA}$ of $J/\psi$ in Pb+Pb collision at LHC as well. Because the collision energy is much higher than at RHIC, parameters will have to be changed from those used before. We change only some of them. $r_o$ and $C$ for Au are set at $6.62~$ fm and $0.546~$ fm respectively \cite{De Jager:1987qc}. $6.4~\mu b$ and $639~\mu b$ are use for the cross sections for $J/\psi$ production and for $c\bar{c}$ pair production per rapidity in p+p collision \cite{Andronic:2006ky}. The constant $30.3$ for produced entropy in Eq. (\ref{entropy}) is replaced by $55.7$ \cite{Andronic:2006ky}. The variable $x$ in Eq. (\ref{entropy}) will be larger than $0.11$, because $x$ increases as collision energy increases. However we use the same value for lack of information. All  chemical potentials are set at zero, because midrapidity region will almost  be baryon-free at LHC energy.
All other parameters including transverse acceleration and its terminal velocity of hot nuclear matter are set at the same values as at RHIC for simplicity and for lack of knowledge.

The lower figure of Fig. \ref{others} shows the prediction for $R_{AA}$ at LHC energy.  In contrast to the case at RHIC energy, the survival rate from thermal decay is very low and the  regeneration effect is dominant except in the peripheral collision. This is so because the lifetime of quark-gluon plasma is much longer than at RHIC. According to our rough estimation, the lifetime at the most central collision is twice that of the RHIC case.
The long lifetime of quark-gluon plasma suppresses the survival rate of charmonia through thermal decay and enhances the relaxation factor $R$ of charm and anti-charm quarks. Abundance of charm quarks is another reason for the dominance of the  regeneration effect.

The same method seems to be applicable to lower collision energies such as SPS. In this case, however, the baryon chemical potential is too high to make use of effective thermal masses by Levai and Heniz, which are extracted from the lattice QCD data with zero baryon chemical potential.

\section{conclusion}

We introduce a generalized two-component model to calculate the $R_{AA}$ of $J/\psi$ near midrapidity in Au+Au collision at $\sqrt{s}=200$ GeV.  In the model, suppression is caused by the nuclear absorption of primordial charmonia and by the thermal decay of charmonia in the  quark-gluon plasma and in the hadron gas.  Enhancement is caused by the regeneration from the QGP. One of our emphases is the use of a  consistent perturbative QCD approach, up to next to the leading order, to estimate the thermal decay widths of charmonia both in the quark-gluon plasma and in the hadron gas.  While the coupling is considered to be a free parameter to be determined by experiment, it is also  related to the screening mass and subsequently dissociation temperatures of charmonia in screened Cornell potential model.

Another  emphasis of our approach  is the initial temperature profile which leads to initial suppression of charmonia if the local temperature is higher than the dissociation temperature.  When the initial temperature profiles, and thus the initial hot regions are considered, the calculated $R_{AA}$ seems to show two sudden drops; one  at very small number of participants and the other at around $170\sim 210$.
The first sudden drop of $R_{AA}$ at very small number of participants is caused when the maximum temperature of hot nuclear matter reaches above the dissociation temperatures of excited charmonia and the second sudden drop when it reaches the   dissociation temperature of $J/\psi$.
On the other hand, if the thermal width of $J/\psi$ is large compared to the inverse time scale of the QGP, the second sudden drop of $R_{AA}$ disappears.
Moreover, some effects can cancel other effects.  For example, a increased thermal width of $J/\psi$ will reduce the $R_{AA}$, while a decreased mass of $J/\psi$ will enhance it.

The present experimental data from RHIC are still large to discriminate all the different and competing mechanisms discussed here.  However, the central values seem to be consistent with the two drops.
In this sense, the maximum temperature of SPS seems to be lower than the melting temperature of $J/\psi$, because the second drop is not seen in $R_{AA}$ there \cite{Quintans:2006wc}.  Similar conclusion could also be made from the data for the Cu+Cu collision at $\sqrt{s}=200$ GeV.
However refined data  for the Au+Au collision at $\sqrt{s}=200$ GeV are required to make a definite conclusion.  Moreover fitting a single data can not disentangle all the different suppression mechanisms separately.
In that sense, the $R_{AA}$ of $J/\psi$ at LHC energy should provide a very interesting information.  As shown in our rough estimate,
$R_{AA}$ at LHC is very different from that at RHIC, in that it is dominated by the regenerated  $J/\psi$ from the QGP.  Moreover, within the thermal decay width given in the paper, the life time of the plasma seems to be much longer than the inverse of the thermal decay with.  Hence, initially formed $J/\psi$ will be mostly suppressed through  thermal decay.
Therefore, the upcoming LHC experiment will provide a totally new information about $J/\psi$ in relativistic heavy ion collision and thus the nature of QGP and its formation.

\bigskip
\section*{Acknowledgements}
This work was supported by
KOSEF(R01-2006-000-10684-0) and by KRF-2006-C00011.  We thank K.Morita for useful discussions.

%%%%%%%%%%%%%%%%%%%%%%%%%%%%%%%%%% Appendix %%%%%%%%%%%%%%%%%%%%%%%%%%%%%%%%%%%%%%%%%%%%%%%%

\hfil\break
\appendix
\centerline{\bf \large Appendix }
\bigskip

\begin{figure}
\centerline{
\includegraphics[width=4 cm, angle=359]{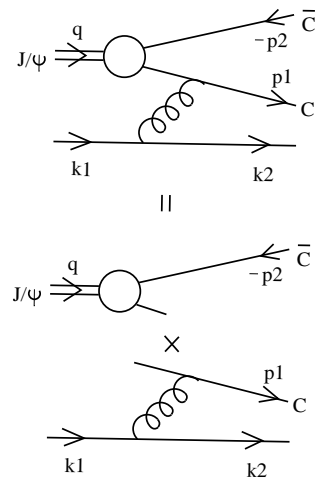}}
\caption{one diagram of quark-induced $J/\psi$ dissociation at next to the leading order in perturbative QCD, and its decomposition}
\label{NLOq}
\end{figure}

The upper figure of Fig. \ref{NLOq} is a diagram for the $J/\psi$ dissociation by a light quark at next to the leading order in perturbative QCD. The double line represents external line of $J/\psi$ and the adjacent small circle the Bethe-Salpeter amplitude. The Bethe-Salpeter vertex represents bound state of charm quark and anti-charm quark and has the following form in the heavy quark limit and in rest frame of $J/\psi$:

\begin{eqnarray}
\Gamma_\mu(p_1,-p_2)&=&\sqrt{\frac{m_{J/\psi}}{N_c}}\bigg(\epsilon_o+\frac{p^2}{m_c}\bigg)\psi(p)\nonumber\\
&&\times \frac{1+\gamma^0}{2}\gamma_i g^i_\mu \frac{1-\gamma^0}{2},
\end{eqnarray}
where $\mu$ is the spin index of $J/\psi$ and $\epsilon_o$ its binding energy. $p$ is the  relative three momentum of $p_1$ and $p_2$, $p\equiv |\vec{p_1}-\vec{p_2}|/2$, and $\psi(p)$ the wavefunction of $J/\psi$. $N_c$ is number of color.
The charm quark propagator between Bethe-Salpeter amplitude and gluon vertex is reexpressed as

\begin{equation}
i\frac{\not{k}+m_c}{k^2-m_c^2}=i\frac{\sum_s u^s(k)\bar{u}^s (k)}{k^2-m_c^2}.
\label{splitting}
\end{equation}

If the binding energy of $J/\psi$ is small, the momentum $k$ of charm quark propagator will be slightly off-shell.
By using Eq. (\ref{splitting}), the invariant amplitude for Fig. \ref{NLOq} is

\begin{eqnarray}
M_\mu &=& \sum_s \bar{u}(k_2)\gamma_\nu u(k_1)\frac{i}{(k_1-k_2)^2-m_c^2}\bar{u}(p_1)\gamma_\nu u^s(k)\nonumber\\
&& \times \bar{u}^s(k)\Gamma_\mu(k,p_2) v(p_2)\frac{i}{k^2-m_c^2}\nonumber\\
&\equiv &\sum_s M_{QF}^s \bar{u}^s(k)\Gamma_\mu(k,-p_2) v(p_2)\frac{i}{k^2-m_c^2}, \label{eq-mqg}
\end{eqnarray}
where $M_{QF}^s$ is the invariant amplitude for $c+q \rightarrow c+q$ with spin $s$ of incoming charm quark in the approximation of quasifree particle \cite{Grandchamp:2002wp}. The lower figure of Fig. \ref{NLOq} shows the diagrammatic decomposition of Eq.~(\ref{eq-mqg}).
In heavy quark limit, the condition of energy conservation,

\begin{eqnarray}
m_{J/\psi}+k_1^0=2m_c+\frac{|\vec{p_1}|^2}{2m_c^2}+\frac{|\vec{p_2}|^2}{2m_c^2}+k_2^0\nonumber
\end{eqnarray}
or
\begin{eqnarray}
k_1^0-k_2^0-\epsilon_o=\frac{|\vec{p_1}|^2}{2m_c^2}+\frac{|\vec{p_2}|^2}{2m_c^2},\nonumber
\end{eqnarray}

gives the following order counting
\begin{equation}
k_1^0 \sim k_2^0 \sim O(mg^4)~~~~~|\vec{p_1}|\sim |\vec{p_2}|\sim O(mg^2),
\label{order-counting}
\end{equation}
because the binding energy $\epsilon_o$ is of $mg^4$ order \cite{Song:2005yd}.
Under this order counting, the denominator of charm quark propagator can be approximated as follows,
\begin{eqnarray}
k^2-m_c^2&=&(p_1-k_1+k_2)^2-m_c^2\nonumber\\
&\approx& -2p_1 \cdot (k_1-k_2)\approx -2m_c(k_1^0-k_2^0)\nonumber\\
&=& -2m_c\bigg(\epsilon_o+\frac{|\vec{p_1}|^2}{2m_c^2}+\frac{|\vec{p_2}|^2}{2m_c^2}\bigg)\nonumber\\
&\approx& -2m_c\bigg(\epsilon_o+\frac{|\vec{p}|^2}{m_c^2}\bigg), \label{app-pro}
\end{eqnarray}
where
\begin{eqnarray}
\vec{p}&=&(\vec{p_1}-\vec{p_2})/2=\vec{p_1}-(\vec{k_1}-\vec{k_2})/2 \approx \vec{p_1},\nonumber\\
\vec{p}&=&(\vec{p_1}-\vec{p_2})/2=-\vec{p_2}+(\vec{k_1}-\vec{k_2})/2 \approx -\vec{p_2}\nonumber
\label{approx}
\end{eqnarray}
are used for the last approximation in Eq.~(\ref{app-pro}). The denominator of charm quark propagator indicates how off-shell the propagator is. We can see the off-shellness is canceled by the same factor in Bethe-Salpeter amplitude. Therefore the invariant amplitude for $J/\psi$ dissociation is

\begin{eqnarray}
M_\mu &=& -2i\ m_c\sqrt{\frac{m_{J/\psi}}{N_c}}\psi(p)\sum_s M_{QF}^s \nonumber\\
&&\times \bar{u}^s(k)\frac{1+\gamma^0}{2}\gamma_i g^i_\mu \frac{1-\gamma^0}{2} v(p_2),
\end{eqnarray}
and its averaged square with respect to the spin and color of incoming particles is

\begin{eqnarray}
|\overline{M}|^2&=&\frac{m_{J/\psi}}{2N_c^2}|\psi(p)|^2|M_{QF}|^2\nonumber\\
&=& 2m_{J/\psi}|\psi(p)|^2 |\overline{M_{QF}}|^2.
\end{eqnarray}
%where is used the following:
%\begin{eqnarray}
%\sum_{s s'}u^{s'}\bar{u}^s M_{QF}^s M_{QF}^{s'\dagger}&=&\sum_{s s'}u^{s'}\bar{u}^s \delta_{ss'}M_{QF}^s %M_{QF}^{s\dagger}\nonumber\\
%&=&\frac{1}{2}\sum_{s'}u^{s'}\bar{u}^{s'}\sum_s M_{QF}^s M_{QF}^{s\dagger}.\nonumber
%\end{eqnarray}}

The phase space of final state is separated as follows,
\begin{eqnarray}
&&\int\frac{d^3p_1 d^3p_2 d^3k_2}{(2\pi)^3 2E_1 (2\pi)^3 2E_2 (2\pi)^3 2k_2^0}\nonumber\\
&&~~~~\times (2\pi)^4 \delta^4(p_1+p_2+k_2-q-k_1)\nonumber\\
&&=\int\frac{d^3p_2 d^4k}{(2\pi)^3 2E_2}\delta^4(k+p_2-q)\nonumber\\
&&\times \int\frac{d^3p_1 d^3k_2}{(2\pi)^3 2E_1 (2\pi)^3 2k_2^0}(2\pi)^4 \delta^4(p_1+k_2-k-k_1).\nonumber\\
\end{eqnarray}

Finally, the dissociation cross section of $J/\psi$ from the diagram is
\begin{eqnarray}
\sigma^{diss}&=&\frac{1}{4m_{J/\psi}k_1^0}\int\frac{d^3p_1 d^3p_2 d^3k_2}{(2\pi)^3 2E_1 (2\pi)^3 2E_2 (2\pi)^3 2k_2^0}\nonumber\\
&&\times (2\pi)^4 \delta^4(p_1+p_2+k_2-q-k_1)|\overline{M}|^2\nonumber\\
&=&\int\frac{d^3p_2 d^4k}{(2\pi)^3 2E_2}\delta^4(k+p_2-q)|\psi(p)|^2\nonumber\\
&&\times \frac{1}{2k_1^0}\int\frac{d^3p_1 d^3k_2}{(2\pi)^3 2E_1 (2\pi)^3 2k_2^0}(2\pi)^4 \nonumber\\
&&\times \delta^4(p_1+k_2-k-k_1)|\overline{M_{QF}}|^2.
\end{eqnarray}

In heavy quark limit,
\begin{equation}
E_2\approx m_c,~~~~k=q-p_2\approx (m_c,\vec{0}),\nonumber
\end{equation}

the dissociation cross section is simplified as follows,
\begin{eqnarray}
\sigma^{diss}&\approx& \int\frac{d^3p}{(2\pi)^3}|\psi(p)|^2 \sigma^{QF},
\label{monopole}
\end{eqnarray}
where $p_2$ is approximated to $-p$.
$\sigma^{QF}$ is the cross section of charm quark, which is used for dissociation cross section of $J/\psi$ in the approximation of quasifree particle.  Suppose that the cross section is independent of $p$, which enters in $\sigma^{QF}$ through the momentum carried by the quasifree charm quark,  then the momentum integral with the square of the wave function appearing in Eq.~(\ref{monopole}) just factors out as the wave function normalization, which is unity, and we will have
\begin{eqnarray}
\sigma^{diss}=\sigma^{QF}.
\end{eqnarray}
In this limit, the cross sections will have to be just a constant independent of the incoming energy.

\begin{figure}
\centerline{
\includegraphics[width=8 cm, angle=359]{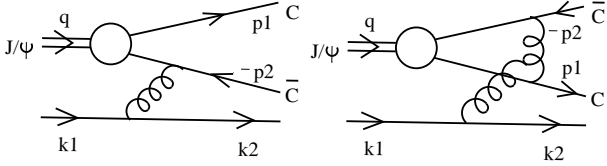}}
\caption{other diagrams of quark-induced $J/\psi$ dissociation at next to the leading order in perturbative QCD}
\label{NLOq2}
\end{figure}

However, Fig. \ref{NLOq} is not the only diagram for quark-induced $J/\psi$ dissociation. The left figure of Fig. \ref{NLOq2} is another diagram for the same dissociation. In the quasifree particle approach, this diagram corresponds to the process $\bar{c}+q\rightarrow \bar{c}+q$ while the diagram of Fig. \ref{NLOq} to the process $c+q\rightarrow c+q$. Moreover, the invariant amplitude for each process is squared and then summed to get dissociation cross section. As a result, cross section becomes double the single contribution.  On the other hand, in our perturbative QCD approach, the amplitude of the two diagrams  are summed before being squared. In this case, the invariant amplitude for the left figure of Fig. \ref{NLOq2} cancels that of Fig. \ref{NLOq} exactly so that the next to the leading order in the counting scheme of Eq. (\ref{order-counting}) has to be considered.
In the order counting, the next to the leading order of Fig. \ref{NLOq}, the left figure of Fig. \ref{NLOq2} and the leading order of the right figure of \ref{NLOq2} are all the same so they are treated equally. It gives the following result \cite{Song:2005yd}:

\begin{eqnarray}
|\overline{M}|^2&=&\frac{4}{3}g^4 m_c^2 m_{J/\psi}\bigg|\frac{\partial\psi(p)}{\partial p}\bigg|^2\nonumber\\
&&\times \bigg\{-\frac{1}{2}+\frac{(k_1^0)^2+(k_2^0)^2}{2k_1\cdot k_2}\bigg\}.
\label{dipole}
\end{eqnarray}

It is worthy to notice that Eq. (\ref{dipole}) is a function of $\partial \psi(p)/\partial p$, the dipole of wavefunction, while Eq. (\ref{monopole}) is a function of $\psi(p)$, the monopole of wavefunction. The result with the dipole form is natural result in the sense that $J/\psi$ is a dipole of color charge.

The comparison between two models in the gluon-induced dissociation cross section is similar. The only difference is that perturbative QCD approach has diagrams where one gluon is absorbed at a heavy quark or antiquark line and the other gluon is emitted at the other heavy antiquark or quark line as shown in diagrams (3) and (4) of Fig. 6 in \cite{Song:2005yd}, but the approach of quasifree particle doesn't have such diagrams.
In addition to the differences discussed above, comparing the quasifree model with ours, one finds that the perturbative QCD approach has contributions from the diagrams where gluon is emitted or absorbed at the gluon line exchanged between the heavy quark and antiquark as shown in (9) $\sim$ (16) of Fig. 6 in \cite{Song:2005yd}.

%%%%%%%%%%%%%%%%%%%%%%%%%%%%%%%%%%%%%%%%%% reference %%%%%%%%%%%%%%%%%%%%%%%%%%%%%%%%%%%%%%%%%%%%%%%%%%%%%%%%%%%%%%%%%


\begin{thebibliography}{10}

%\cite{Matsui:1986dk} 1
\bibitem{Matsui:1986dk}
  T.~Matsui and H.~Satz,
  %``J/psi Suppression by Quark-Gluon Plasma Formation,''
  Phys.\ Lett.\  B {\bf 178}, 416 (1986).
  %%CITATION = PHLTA,B178,416;%%


\bibitem{Hatsuda04}
  M.~Asakawa and T.~Hatsuda,
  %``J/psi and eta/c in the deconfined plasma from lattice QCD,''
  Phys. Rev. Lett. {\bf 92}, 012001 (2004)
  %[arXiv:hep-lat/0308034].
  %%CITATION = HEP-LAT 0308034;%%

\bibitem{Datta04}
  S.~Datta, F.~Karsch, P.~Petreczky and I.~Wetzorke,
  %``Behavior of charmonium systems after deconfinement,''
  Phys. Rev. D {\bf 69}, 094507 (2004)
  %[arXiv:hep-lat/0312037].
  %%CITATION = HEP-LAT 0312037;%%

%\cite{Mocsy:2007jz} 10
\bibitem{Mocsy:2007jz}
  A.~Mocsy and P.~Petreczky,
  %``Color Screening Melts Quarkonium,''
  Phys.\ Rev.\ Lett.\  {\bf 99}, 211602 (2007)
  [arXiv:0706.2183 [hep-ph]].
  %%CITATION = PRLTA,99,211602;%%

\bibitem{Morita:2007pt}
  K.~Morita and S.~H.~Lee,
  %``Mass shift and width broadening of J/psi in QGP from QCD sum rule,''
  Phys.\ Rev.\ Lett.\  {\bf 100}, 022301 (2008)
  [arXiv:0704.2021 [nucl-th]].
  %%CITATION = PRLTA,100,022301;%%

%\cite{Lee:2008xp} 30
\bibitem{Lee:2008xp}
  S.~H.~Lee and K.~Morita,
  %``Properties of $J/\psi$ at $T_c$: QCD second-order Stark effect,''
  Phys.\ Rev.\  D {\bf 79}, 011501 (2009)
  [arXiv:0802.4000 [hep-ph]].
  %%CITATION = PHRVA,D79,011501;%%

\bibitem{Gazdzicki:1999rk}
  M.~Gazdzicki and M.~I.~Gorenstein,
  %``Evidence for statistical production of $J/\psi$ mesons in nuclear
  %collisions at the CERN SPS,''
  Phys.\ Rev.\ Lett.\  {\bf 83}, 4009 (1999)
  [arXiv:hep-ph/9905515].
  %%CITATION = PRLTA,83,4009;%%

%\cite{BraunMunzinger:2000px} 23
\bibitem{BraunMunzinger:2000px}
  P.~Braun-Munzinger and J.~Stachel,
  %``(Non)thermal aspects of charmonium production and a new look at J/psi
  %suppression,''
  Phys.\ Lett.\  B {\bf 490}, 196 (2000)
  [arXiv:nucl-th/0007059].
  %%CITATION = PHLTA,B490,196;%%

\bibitem{Gorenstein:2000ck}
  M.~I.~Gorenstein, A.~P.~Kostyuk, H.~Stoecker and W.~Greiner,
  %``Statistical coalescence model with exact charm conservation,''
  Phys.\ Lett.\  B {\bf 509}, 277 (2001)
  [arXiv:hep-ph/0010148].
  %%CITATION = PHLTA,B509,277;%%

\bibitem{Thews:2000rj}
  R.~L.~Thews, M.~Schroedter and J.~Rafelski,
  %``Enhanced J/psi production in deconfined quark matter,''
  Phys.\ Rev.\  C {\bf 63}, 054905 (2001)
  [arXiv:hep-ph/0007323].
  %%CITATION = PHRVA,C63,054905;%%


%\cite{Back:2004je} 2
\bibitem{Back:2004je}
  B.~B.~Back {\it et al.},
  %``The PHOBOS perspective on discoveries at RHIC,''
  Nucl.\ Phys.\  A {\bf 757}, 28 (2005)
  [arXiv:nucl-ex/0410022].
  %%CITATION = NUPHA,A757,28;%%


%\cite{Adare:2006ns} 3
\bibitem{Adare:2006ns}
  A.~Adare {\it et al.}  [PHENIX Collaboration],
  %``J/psi production vs centrality, transverse momentum, and rapidity in Au  +
  %Au collisions at s(NN)**(1/2) = 200-GeV,''
  Phys.\ Rev.\ Lett.\  {\bf 98}, 232301 (2007)
  [arXiv:nucl-ex/0611020].
  %%CITATION = PRLTA,98,232301;%%


%\cite{Andronic:2006ky} 4
\bibitem{Andronic:2006ky}
  A.~Andronic, P.~Braun-Munzinger, K.~Redlich and J.~Stachel,
  %``Statistical hadronization of heavy quarks in ultra-relativistic
  %nucleus-nucleus collisions,''
  Nucl.\ Phys.\  A {\bf 789}, 334 (2007)
  [arXiv:nucl-th/0611023].
  %%CITATION = NUPHA,A789,334;%%


%\cite{Grandchamp:2002wp} 26
\bibitem{Grandchamp:2002wp}
  L.~Grandchamp and R.~Rapp,
  %``Charmonium suppression and regeneration from SPS to RHIC,''
  Nucl.\ Phys.\  A {\bf 709}, 415 (2002)
  [arXiv:hep-ph/0205305].
  %%CITATION = NUPHA,A709,415;%%

%\cite{Zhao:2007hh} 5
\bibitem{Zhao:2007hh}
  X.~Zhao and R.~Rapp,
  %``Transverse Momentum Spectra of J/\psi in Heavy-Ion Collisions,''
  Phys.\ Lett.\  B {\bf 664}, 253 (2008)
  [arXiv:0712.2407 [hep-ph]].
  %%CITATION = PHLTA,B664,253;%%


%\cite{Yan:2006ve} 6
\bibitem{Yan:2006ve}
  L.~Yan, P.~Zhuang and N.~Xu,
  %``Competition between J/psi suppression and regeneration in quark-gluon
  %plasma,''
  Phys.\ Rev.\ Lett.\  {\bf 97}, 232301 (2006)
  [arXiv:nucl-th/0608010].
  %%CITATION = PRLTA,97,232301;%%


\bibitem{Wong04}
  C.~Y.~Wong,
  %``Heavy quarkonia in quark gluon plasma,''
  Phys.\ Rev.\ C {\bf 72}, 034906 (2005)
  [arXiv:hep-ph/0408020].
  %%CITATION = HEP-PH 0408020;%%


%\cite{Petreczky:2010yn}
\bibitem{Petreczky:2010yn}
  P.~Petreczky,
  %``Quarkonium in Hot Medium,''
  arXiv:1001.5284 [hep-ph].
  %%CITATION = ARXIV:1001.5284;%%

%\cite{Levai:1997yx} 11
\bibitem{Levai:1997yx}
  P.~Levai and U.~W.~Heinz,
  %``Massive gluons and quarks and the equation of state obtained from SU(3)
  %lattice QCD,''
  Phys.\ Rev.\  C {\bf 57}, 1879 (1998)
  [arXiv:hep-ph/9710463].
  %%CITATION = PHRVA,C57,1879;%%


%\cite{BraunMunzinger:1999qy}
%\bibitem{BraunMunzinger:1999qy}
%  P.~Braun-Munzinger, I.~Heppe and J.~Stachel,
  %``Chemical equilibration in Pb + Pb collisions at the SPS,''
%  Phys.\ Lett.\  B {\bf 465}, 15 (1999)
%  [arXiv:nucl-th/9903010].
  %%CITATION = PHLTA,B465,15;%%


%\cite{Antinori:2000ph} 12
\bibitem{Antinori:2000ph}
  F.~Antinori {\it et al.}  [WA97 Collaboration and NA57 Collaborations],
  %``Determination of the number of wounded nucleons in Pb + Pb collisions  at
  %158-A-GeV/c,''
  Nucl.\ Phys.\  A {\bf 661}, 357 (1999)
  [Eur.\ Phys.\ J.\  C {\bf 18}, 57 (2000)].
  %%CITATION = EPHJA,C18,57;%%


%\cite{De Jager:1987qc} 13
\bibitem{De Jager:1987qc}
  H.~De Vries, C.~W.~De Jager and C.~De Vries,
  %``Nuclear charge and magnetization density distribution parameters from
  %elastic electron scattering,''
  Atom.\ Data Nucl.\ Data Tabl.\  {\bf 36}, 495 (1987).
  %%CITATION = ADNDA,36,495;%%


%\cite{Adcox:2004mh}
%\bibitem{Adcox:2004mh}
%  K.~Adcox {\it et al.}  [PHENIX Collaboration],
  %``Formation of dense partonic matter in relativistic nucleus nucleus
  %collisions at RHIC: Experimental evaluation by the PHENIX  collaboration,''
%  Nucl.\ Phys.\  A {\bf 757}, 184 (2005)
%  [arXiv:nucl-ex/0410003].
  %%CITATION = NUPHA,A757,184;%%


\bibitem{Morita02}
  K.~Morita, S.~Muroya, C.~Nonaka and T.~Hirano,
  %``Comparison of space-time evolutions of hot/dense matter in  s(NN)**(1/2) =
  %17-GeV and 130-GeV relativistic heavy ion collisions  based on a
  %hydrodynamical model,''
  Phys.\ Rev.\  C {\bf 66}, 054904 (2002)
  [arXiv:nucl-th/0205040].
  %%CITATION = PHRVA,C66,054904;%%

%\cite{Hirano:2001eu}
\bibitem{Hirano:2001eu}
  T.~Hirano,
  %``Is early thermalization achieved only near midrapidity in Au + Au
  %collisions at sNN =130 GeV?,''
  Phys.\ Rev.\  C {\bf 65}, 011901 (2002)
  [arXiv:nucl-th/0108004].
  %%CITATION = PHRVA,C65,011901;%%

%\cite{Gunji:2007uy}
\bibitem{Gunji:2007uy}
  T.~Gunji, H.~Hamagaki, T.~Hatsuda and T.~Hirano,
  %``Onset of J/psi melting in quark-gluon fluid at RHIC,''
  Phys.\ Rev.\  C {\bf 76}, 051901 (2007)
  [arXiv:hep-ph/0703061].
  %%CITATION = PHRVA,C76,051901;%%


%\cite{Kharzeev:2000ph}
\bibitem{Kharzeev:2000ph}
  D.~Kharzeev and M.~Nardi,
  %``Hadron production in nuclear collisions at RHIC and high density QCD,''
  Phys.\ Lett.\  B {\bf 507}, 121 (2001)
  [arXiv:nucl-th/0012025].
  %%CITATION = PHLTA,B507,121;%%


%\cite{Back:2002uc}
\bibitem{Back:2002uc}
  B.~B.~Back {\it et al.}  [PHOBOS Collaboration],
  %``Centrality dependence of the charged particle multiplicity near
  %mid-rapidity in Au + Au collisions at $\sqrt{s}$ (NN) = 130-GeV and
  %200-GeV,''
  Phys.\ Rev.\  C {\bf 65}, 061901 (2002)
  [arXiv:nucl-ex/0201005].
  %%CITATION = PHRVA,C65,061901;%%


%\cite{Schneider:2001nf}
\bibitem{Schneider:2001nf}
  R.~A.~Schneider and W.~Weise,
  %``On the quasiparticle description of lattice QCD thermodynamics,''
  Phys.\ Rev.\  C {\bf 64}, 055201 (2001)
  [arXiv:hep-ph/0105242].
  %%CITATION = PHRVA,C64,055201;%%


%\cite{Song:2007gm} 14
\bibitem{Song:2007gm}
  T.~Song, Y.~Park, S.~H.~Lee and C.~Y.~Wong,
  %``The thermal width of heavy quarkonia moving in quark gluon plasma,''
  Phys.\ Lett.\  B {\bf 659}, 621 (2008)
  [arXiv:0709.0794 [hep-ph]].
  %%CITATION = PHLTA,B659,621;%%


%\cite{Park:2007zza} 15
\bibitem{Park:2007zza}
  Y.~Park, K.~I.~Kim, T.~Song, S.~H.~Lee and C.~Y.~Wong,
  %``Widths of quarkonia in quark gluon plasma,''
  Phys.\ Rev.\  C {\bf 76}, 044907 (2007)
  [arXiv:0704.3770 [hep-ph]].
  %%CITATION = PHRVA,C76,044907;%%


%\cite{Song:2005yd} 16
\bibitem{Song:2005yd}
  T.~Song and S.~H.~Lee,
  %``Quarkonium hadron interactions in perturbative QCD,''
  Phys.\ Rev.\  D {\bf 72}, 034002 (2005)
  [arXiv:hep-ph/0501252].
  %%CITATION = PHRVA,D72,034002;%%


%\cite{Karsch:1987pv} 21
\bibitem{Karsch:1987pv}
  F.~Karsch, M.~T.~Mehr and H.~Satz,
  %``Color Screening and Deconfinement for Bound States of Heavy Quarks,''
  Z.\ Phys.\  C {\bf 37}, 617 (1988).
  %%CITATION = ZEPYA,C37,617;%%


%\cite{Peskin:1979va}
\bibitem{Peskin:1979va}
  M.~E.~Peskin,
  %``Short Distance Analysis For Heavy Quark Systems. 1. Diagrammatics,''
  Nucl.\ Phys.\  B {\bf 156}, 365 (1979).
  %%CITATION = NUPHA,B156,365;%%


%\cite{Bhanot:1979vb}
\bibitem{Bhanot:1979vb}
  G.~Bhanot and M.~E.~Peskin,
  %``Short Distance Analysis For Heavy Quark Systems. 2. Applications,''
  Nucl.\ Phys.\  B {\bf 156}, 391 (1979).
  %%CITATION = NUPHA,B156,391;%%


%\cite{Sutton:1991ay} 17
\bibitem{Sutton:1991ay}
  P.~J.~Sutton, A.~D.~Martin, R.~G.~Roberts and W.~J.~Stirling,
  %``Parton distributions for the pion extracted from Drell-Yan and prompt
  %photon experiments,''
  Phys.\ Rev.\  D {\bf 45}, 2349 (1992).
  %%CITATION = PHRVA,D45,2349;%%


%\cite{Lin:1999ad} 18
\bibitem{Lin:1999ad}
  Z.~w.~Lin and C.~M.~Ko,
  %``Model for $J/\psi$ absorption in hadronic matter,''
  Phys.\ Rev.\  C {\bf 62}, 034903 (2000)
  [arXiv:nucl-th/9912046].
  %%CITATION = PHRVA,C62,034903;%%


%\cite{Oh:2000qr} 19
\bibitem{Oh:2000qr}
  Y.~s.~Oh, T.~Song and S.~H.~Lee,
  %``J/psi absorption by pi and rho mesons in meson exchange model with
  %anomalous parity interactions,''
  Phys.\ Rev.\  C {\bf 63}, 034901 (2001)
  [arXiv:nucl-th/0010064].
  %%CITATION = PHRVA,C63,034901;%%


%\cite{Wong:1999zb} 20
\bibitem{Wong:1999zb}
  C.~Y.~Wong, E.~S.~Swanson and T.~Barnes,
  %``Cross sections for pi and rho induced dissociation of J/psi and psi',''
  Phys.\ Rev.\  C {\bf 62}, 045201 (2000)
  [arXiv:hep-ph/9912431].
  %%CITATION = PHRVA,C62,045201;%%



%\cite{LindenLevy:2009zz} 22
\bibitem{LindenLevy:2009zz}
  L.~A.~Linden Levy  [PHENIX Collaboration],
  %``Disentangling charmonium suppression,''
  J.\ Phys.\ G {\bf 36}, 064013 (2009).
  %%CITATION = JPHGB,G36,064013;%%





%\cite{Cacciari:2005rk} 24
\bibitem{Cacciari:2005rk}
  M.~Cacciari, P.~Nason and R.~Vogt,
  %``QCD predictions for charm and bottom production at RHIC,''
  Phys.\ Rev.\ Lett.\  {\bf 95}, 122001 (2005)
  [arXiv:hep-ph/0502203].
  %%CITATION = PRLTA,95,122001;%%


%\cite{Combridge:1978kx} 27
\bibitem{Combridge:1978kx}
  B.~L.~Combridge,
  %``Associated Production Of Heavy Flavor States In P P And Anti-P P
  %Interactions: Some QCD Estimates,''
  Nucl.\ Phys.\  B {\bf 151}, 429 (1979).
  %%CITATION = NUPHA,B151,429;%%


%\cite{Adare:2006kf} 28
\bibitem{Adare:2006kf}
  A.~Adare {\it et al.}  [PHENIX Collaboration],
  %``$J/\psi$ production versus transverse momentum and rapidity in $p^+ p$
  %collisions at $\sqrt{s}$ = 200-GeV,''
  Phys.\ Rev.\ Lett.\  {\bf 98}, 232002 (2007)
  [arXiv:hep-ex/0611020].
  %%CITATION = PRLTA,98,232002;%%

\bibitem{Kaczmarek}
  O.~Kaczmarek and F.~Zantow,
  %``Static quark anti-quark interactions in zero and finite temperature  QCD.
  %I: Heavy quark free energies, running coupling and quarkonium  binding,''
  Phys.\ Rev.\  D {\bf 71}, 114510 (2005)
  [arXiv:hep-lat/0503017].
  %%CITATION = PHRVA,D71,114510;%%


%\cite{Song:2008bd} 31
\bibitem{Song:2008bd}
  Y.~H.~Song, S.~H.~Lee and K.~Morita,
  %``In-medium modification of P-wave charmonia from QCD sum rules,''
  Phys.\ Rev.\  C {\bf 79}, 014907 (2009)
  [arXiv:0808.1153 [hep-ph]].
  %%CITATION = PHRVA,C79,014907;%%



%\cite{Quintans:2006wc} 29
\bibitem{Quintans:2006wc}
  C.~Quintans {\it et al.}  [NA50 Collaboration],
  %``Na50 Final Results On Charmonia Suppression,''
  J.\ Phys.\ Conf.\ Ser.\  {\bf 50}, 353 (2006).
  %%CITATION = 00462,50,353;%%


\end{thebibliography}
\end{document}